\documentclass[12pt,twoside]{article}
\usepackage{a4}
\usepackage{a4wide}
\usepackage{ulheadings}
\usepackage{epsfig}
\usepackage[latin1]{inputenc}
\newcommand{\be}{\begin{equation}}
\newcommand{\ee}{\end{equation}}
\newcommand{\ba}{\begin{array}{c}}
\newcommand{\baz}{\begin{array}{cc}}
\newcommand{\bad}{\begin{array}{ccc}}
\newcommand{\bea}{\begin{equation} \begin{array}{c}}
\newcommand{\eea}{ \end{array} \end{equation}}
\newcommand{\ea}{\end{array}}
\newcommand{\D}{\displaystyle}
\newcommand{\en}{\mbox{$\nu_e$}}
\newcommand{\ra}{\mbox{ $\rightarrow$ }}
\newcommand{\mun}{\mbox{$\nu_{\mu}$}}
\newcommand{\taun}{\mbox{$\nu_{\tau}$}}

\newcommand{\nua}{\mbox{$\nu_a$}}

\newcommand{\nual}{\mbox{$\nu_{\alpha}$}}
\newcommand{\nube}{\mbox{$\nu_{\beta}$}}

\newcommand{\lra}{\leftrightarrow}
\newcommand{\Deu}{\mbox{D}}
\newcommand{\mn}{\mbox{$m_{\nu}$}}
\def\ga{\mathrel{
   \rlap{\raise 0.511ex \hbox{$>$}}{\lower 0.511ex \hbox{$\sim$}}}}
\def\sa{\mathrel{
   \rlap{\raise 0.511ex \hbox{$<$}}{\lower 0.511ex \hbox{$\sim$}}}}

\begin{document}


\begin{titlepage}

\begin{center}

{\hfill { \normalsize DO--TH 99/22}\\
\hfill {\normalsize December 1999}\\}
\vskip .5in
{\Large \bf Massive Neutrinos in Astrophysics}\\ [1.1cm]

{\large\bf Lectures at the fourth national summer school\\
``Grundlagen und neue Methoden der theoretischen Physik''\\
\vskip4pt
31 August -- 11 September 1998, Saalburg, Germany}

\vskip .4in

{\large\bf Georg G.\ Raffelt}  
\\
\vskip 0.4cm
{\em Max--Planck--Institut für Physik \\ 
(Werner--Heisenberg--Institut)\\
        Föhringer Ring 6, 80805 München, Germany}\\
\smallskip
{\rm Email:~raffelt@mppmu.mpg.de}
\vskip 0.8cm

Lecture notes written by\\
\bigskip
{\large\bf Werner Rodejohann } 
\\
\vskip 0.4cm
{\em Lehrstuhl Theoretische Physik III\\ 
Universität Dortmund\\
Otto--Hahn Str.\ 4, 44221 Dortmund, Germany}\\
\smallskip
{\rm Email:~rodejoha@dilbert.physik.uni-dortmund.de}
\end{center}

\vskip 1.1cm

\begin{center} {\bf ABSTRACT } \end{center}

\noindent 
An introduction to various topics in neutrino astrophysics is given for
students with little prior exposure to this field.  We explain neutrino
production and propagation in stars, neutrino oscillations, and experimental
searches for this effect.  We also touch upon the cosmological role of
neutrinos.  A number of exercises is also included.



\end{titlepage}

\eject

\newpage

\addcontentsline{toc}{section}{Table of Contents}
\pagenumbering{roman}
\tableofcontents
\newpage


\addcontentsline{toc}{section}{Preface}

\section*{Preface}

Neutrino astrophysics is a prime example for the modern connection between
astrophysics and particle physics which is often referred to as
``astroparticle physics'' or also ``particle astrophysics.'' The intrinsic
properties of neutrinos, especially the question of their mass, is one of the
unsolved problems of particle physics. On the other hand, neutrino masses and
other more hypothetical properties such as electromagnetic couplings can play
an important role in various astrophysical environments. Therefore,
astrophysics plays an important role at constraining nonstandard neutrino
properties.

Of course, the most exciting recent development is the overwhelming evidence
for neutrino oscillations from the atmospheric neutrino anomaly, and notably
the zenith--angle variation observed in the SuperKamiokande experiment.
Besides the near--certainty that the phenomenon of neutrino oscillations is
real, this high--statistics experiment has also opened a new era in neutrino
astronomy. It may not be too long until large--scale neutrino telescopes
observe novel astrophysical sources in the ``light'' of neutrino radiation.

In these lectures we cover a number of topics in the area of neutrino
astrophysics and cosmology which are of current interest to an audience of
students who have not had much prior exposure to either neutrino physics,
astrophysics, or cosmology. At the summer school, the lectures were presented
on a chalk board, with only a small number of viewgraph projections, severely
limiting the amount of material that could be covered in a few hours.  Some of
the material was treated in two exercise sessions; some of the exercises are
integrated into the present notes.  Still, these lectures are rather
incomplete and give only a first impression of the field.

For a more complete coverage the reader is referred to the excellent textbook
by Schmitz~\cite{schmitz}, which unfortunately is available only in German.
Many of the stellar--evolution topics are covered in ``Stars as Laboratories
for Fundamental Physics'' by Raffelt~\cite{raffelt}.  For the cosmological
questions, the best textbook reference remains the classic by Kolb and
Turner~\cite{kolbturner}. A good overview of many of the relevant issues is
provided in two recent textbooks on astroparticle
physics~\cite{zuber,bergstrom}. Finally, we mention a few recent review
articles which may be of help to access the field in more
depth~\cite{zuber2,raffelt98,raffelt99,bilenky99}.

Chapter~1 treats the production of neutrinos in normal stars, especially in
the Sun, but leaving out supernova physics ---there simply was not enough time
to treat this complicated topic in the lectures.  Chapter~2 discusses neutrino
oscillations in vacuum and in matter.  In Chapter~3, experimental strategies
are reviewed and some experiments are described in more detail.  Chapter~4 is
devoted to the connection of neutrinos and cosmology. Conclusions are 
presented in Chpater 5.


\newpage

\pagenumbering{arabic}
\pagestyle{headings}
\section{Neutrinos in Normal Stars}

\subsection{The Sun}

Ever since it became clear that stars are powered by nuclear fusion reactions
and that neutrinos are produced in nuclear reactions, it was also clear that
stars are powerful neutrino sources. Stellar evolution proceeds through many
distinct evolutionary phases~\cite{clayton,kippenhahn}.  Stars spend most of
their lives on the initial ``main--sequence,'' the Sun is an example, where
energy is gained from the fusion of hydrogen to helium, i.e.\ by the net
reaction \be 
\label{sonfus} 4p+2e^- \ra {}^4{\rm He} + 2\en + 26.73~{\rm MeV}.
\ee 
The detailed reaction chains and cycles depend on the stellar mass which,
in turn, influences the equilibrium temperature in the interior.  In the case
of a low--mass star like the Sun, hydrogen burning proceeds primarily through
the $pp$ chains. The CNO cycle (Bethe--Weizsäcker cycle), which dominates in
more massive stars, contributes only about 1.6\%.

\begin{table}[b]
\caption{\label{solarreactions}Solar neutrino production in the $pp$ chains.}
\bigskip
\hbox to\hsize{\hss
\begin{tabular}{llrrr}
\hline
\hline
\noalign{\vskip4pt}
Name&Reaction&\multicolumn{1}{c}{$\langle E_\nu\rangle$}&
\multicolumn{1}{c}{$E_\nu^{\rm max}$}&
\multicolumn{1}{c}{Fractional}\\
&&\multicolumn{1}{c}{[MeV]}&
\multicolumn{1}{c}{[MeV]}&
\multicolumn{1}{c}{solar flux}\\
\noalign{\vskip4pt}
\hline
$pp$ & $p+p \ra {\rm D} + e^+ + \en$&0.26& 0.42 &0.909\\ 
$ pep$ &$p + e^- + p \ra {\rm D} + \en$&1.44&---&$2\times10^{-3}$\\ 
$hep $ &${\rm ^3He} + p \ra {\rm ^4He} + e^+ + \en$&9.62& 18.77 
&$2 \times 10^{-8}$\\
$ {\rm ^7Be}$ &${\rm ^7Be} + e^-\ra  {\rm ^7Li} + \en$&(90\%) 0.86&---&0.074\\
         &                                       &(10\%) 0.38&---&     \\
${\rm ^8B} $ &${\rm ^8B} \ra  {\rm ^8Be}^{\ast} + e^+ + \en$&6.71& $\approx$ 15& 
$8.6 \times 10^{-5}$\\ 
\hline
\end{tabular}\hss}
\end{table}

From the particle physics perspective, the solar neutrino flux is perhaps the
most important example because it has been measured in several different
experiments, giving rise to the ``solar neutrino problem'' and thus provides
evidence for neutrino oscillations.  From Eq.~(\ref{sonfus}) one sees that two
electron neutrinos are produced for every 26.7~MeV of liberated nuclear
energy. Assuming that the neutrinos themselves carry away only a small
fraction of this energy, the total solar flux at Earth
can be estimated as
\be
\Phi_{\en} \simeq 2\,\frac{S}{26.7~{\rm MeV}} =2\, 
\frac{8.5 \times 10^{11}~{\rm MeV~cm^{-2}~s^{-1}}}{26.7~\rm MeV} 
= 6.4 \times  10^{10}~{\rm cm^{-2}~s^{-1}},
\ee
where $S$ is the solar constant. 

\begin{samepage}
\begin{figure}
\setlength{\unitlength}{1cm}
\begin{center}
\vspace{5.2cm}
\epsfig{file=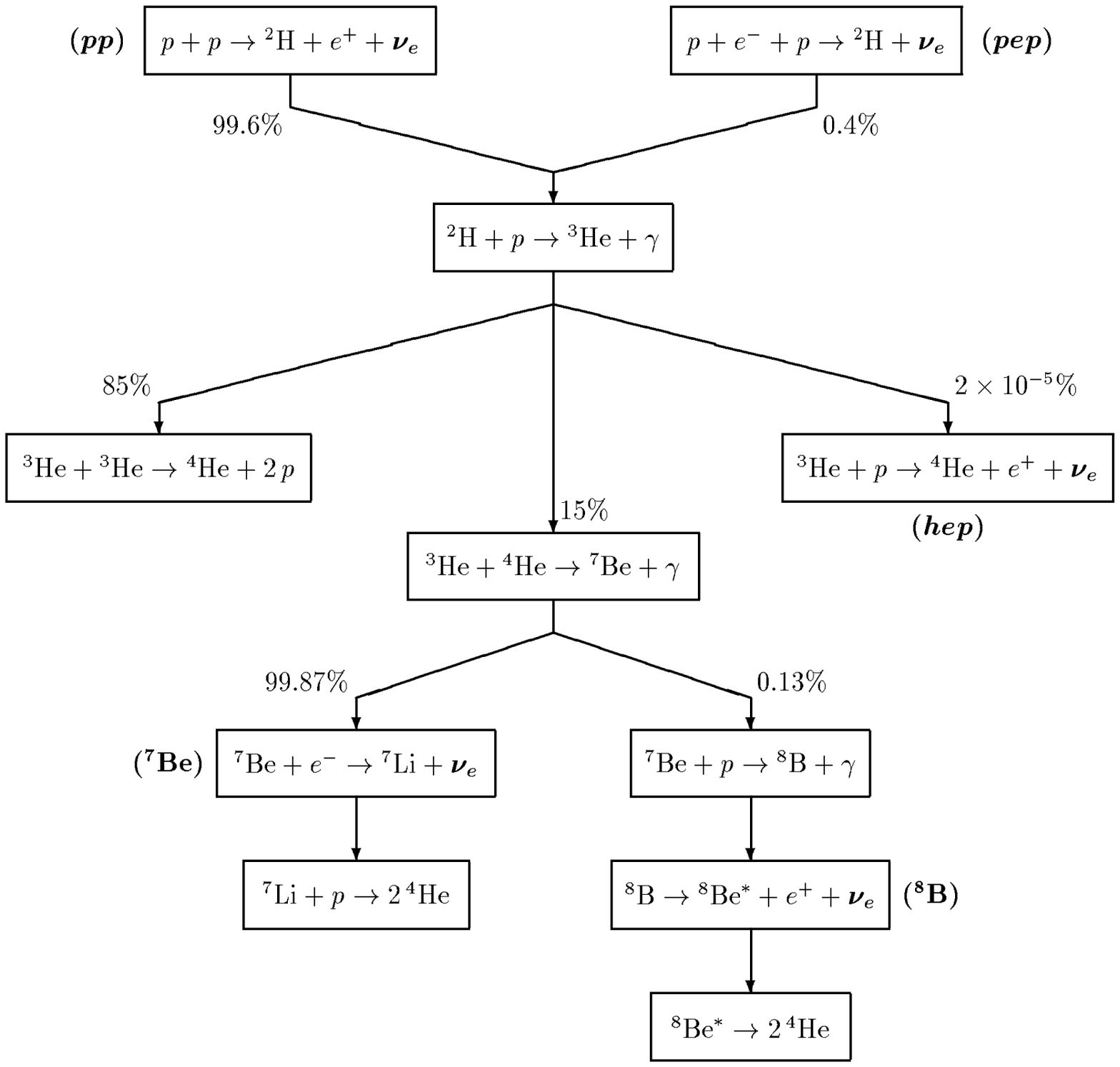,width=12cm}
\end{center}
\vspace{-7cm}
\caption{\label{ppcyclus}Energy generation in the Sun via the $pp$ chains. 
(Figure from Ref.~\cite{bilenky99}.)}
\begin{center}
\epsfig{file=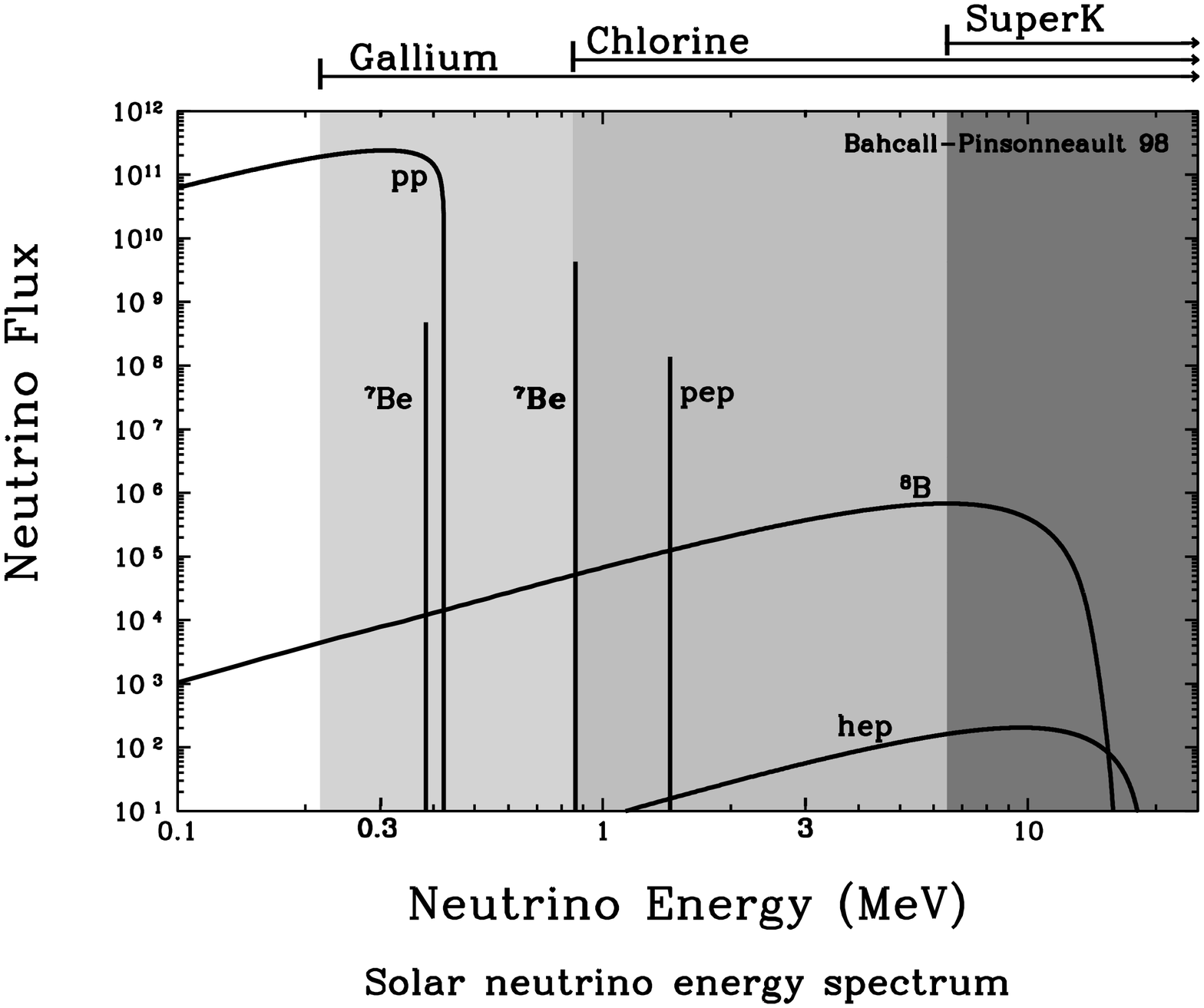,width=8.5cm,angle=270}
\end{center}
\caption{\label{spectrum}Solar neutrino spectrum and thresholds of 
solar neutrino experiments as indicated above the figure
(taken from http://www.sns.ias.edu/$\sim$jnb/).}
\end{figure}
\end{samepage}

However, for the interpretation of the experiments, the detailed spectral
characteristics of the solar neutrino flux are of great importance.  In the
$pp$ chains, electron neutrinos are produced in six different reactions,
giving rise to as many different spectral contributions
(Table~\ref{solarreactions}, Figs.~\ref{ppcyclus} and~\ref{spectrum}), i.e.\ 
three monochromatic lines from electron capture reactions and three continuous
beta--spectra. The sum of the fractional solar neutrino fluxes in
Table~\ref{solarreactions} is less than unity due to the small CNO
contribution.

The detailed contribution of each reaction is based on a standard solar model
(SSM)~\cite{bahcall} which describes the Sun on the basis of a well--defined
set of input assumptions. There is broad consensus in the literature on the
properties of a SSM\@. However, if the real Sun is indeed well represented by a
SSM is not a trivial question. The enormous recent progress in the field of
helioseismology, however, appears to confirm many detailed properties of the
SSM\@. One measures the frequencies of the solar 
pressure modes (p modes) by the
Doppler shift of spectral lines.  One thus probes indirectly the speed of
sound in the Sun at various depths, i.e.\ one can reconstruct a sound--speed
profile of the Sun which is extremely sensitive to the temperature, density
and composition profile.

\subsection{Photon Dispersion in Stellar Plasmas}

The nuclear reactions discussed thus far produce electron neutrinos in beta
reactions where a $\en$ appears together with an $e^+$ or at the expense of
the absorption of an $e^-$.  However, stellar plasmas emit neutrinos also by a
variety of processes where $\nu\bar\nu$ pairs of all flavors appear by
effective neutral--current reactions. The most important cases are
\begin{eqnarray}
&&\hbox to10em{Photo Process\hfil}\gamma+e^-\ra e^-+\nu+\bar\nu,
\nonumber\\
&&\hbox to10em{Pair Annihilation\hfil}e^++e^-\ra \nu+\bar\nu,
\nonumber\\
&&\hbox to10em{Bremsstrahlung\hfil}e^-+(A,Z)\to(A,Z)+e^-+ \nu+\bar\nu,
\nonumber\\
&&\hbox to10em{Plasma Process\hfil}\gamma\to\nu+\bar\nu.
\end{eqnarray}
The relative importance of these processes depends on the temperature and
density of the plasma. The energy of the neutrinos produced in these
reactions is of the order of the temperature of the plasma, 
in contrast with the
nuclear reactions where the neutrino energy is determined by nuclear binding
energies. In the Sun, the central temperature is about 1.3~keV so that thermal
neutrino energies are much smaller than those produced in the $pp$ chains.
The total energy emitted by these processes in the Sun is negligibly small.

However, in later evolutionary phases, neutrinos produced by plasma processes
become much more important than nuclear processes. In particular, the plasma
process (``photon decay'') is the dominant neutrino--producing reaction in
white dwarfs or the cores of horizontal--branch stars or low--mass red
giants. This process is noteworthy because it is not possible in vacuum due
to energy--momentum conservation. In a plasma, on the other hand, the photon
acquires a nontrivial dispersion relation (``effective mass'') so that its
decay into neutrinos is kinematically possible. 

We will see that medium--induced modifications of particle dispersion relations
are not only important for the plasma process, but also for neutrino
oscillations in the Sun or in other environments. One usually defines a
refractive index $n_{\rm refr}$ which relates wave number and frequency of a
particle by $k=n_{\rm refr}\omega$. Refraction in a medium arises from the
interference of the incoming wave with the scattered waves in the forward
direction. Therefore, the refractive index is given in terms of the
forward--scattering amplitude $f_0$ by
\be\label{eq:fowardscattering}
n_{\rm refr} = 1 + \frac{2 \pi}{\omega^2} n f_0 (\omega), 
\ee
where $n$ is the number density of the scattering targets. This formula
applies to any particle propagating in a medium, except that $f_0$ must be
calculated according to the interactions of that particle with the medium
constituents.

Photons interact electromagnetically; the dominant scattering process is
Compton scattering on electrons $\gamma+e^-\ra e^-+\gamma$.  In the low--energy
limit (Thomson scattering) one finds $f_0 = -\alpha/m_e$ with $\alpha=1/137$
the fine--structure constant. It is then trivial to show that the refractive
index corresponds to a dispersion relation
\be\label{gamdisrel}
\omega^2-k^2= \omega_{\rm p}^2=\frac{4 \pi \alpha}{m_e}\, n_e,
\ee 
where $\omega_{\rm p}$ is the so--called plasma frequency.  In the Sun, for
example, one finds $\omega_{\rm p}\simeq0.3~{\rm keV}$ while in the core of
low--mass red giants it is about 9~keV (Exercise~\ref{aufg1}).  The plasma
frequency plays the role of an effective photon mass.  We stress, however,
that the general photon dispersion relation or that of other particles can not
be written in this simple form, i.e.\ in general the effect of dispersion can
not be represented by an effective in--medium mass.

\subsection{\label{muis}Neutrino Refraction in Media}

We next turn to the neutrino dispersion relation in media.  Usually we will be
concerned with very low energies. Therefore, we may take the low--energy limit
of the weak--interaction Hamiltonian where the propagators for the massive
gauge bosons are expanded as
\be
D_{\mu \nu} = \frac{g_{\mu \nu}}{Q^2 - M_{W,Z}^2} 
\simeq\frac{-g_{\mu \nu}}{M_{W,Z}^2}. 
\ee
In this limit one obtains the usual current--current Hamiltonian for the
neutrino--fermion interaction,
\be\label{eq:hamiltonian}
\mbox{$\cal H$}_{\rm int} = \frac{G_F}{\sqrt{2}}\,
\bar{\psi}_f \gamma_{\mu} (C_V - C_A \gamma_5) \psi_{f'} 
\bar{\psi}_{\nu} \gamma^{\mu} (1 - \gamma_5) \psi_{\nu}, 
\ee
where $G_F$ is the Fermi constant.

One may then proceed to calculate the dispersion relation in a medium on the
basis of Eq.~(\ref{eq:fowardscattering}). However, in the special case of a
current--current interaction the neutrino energy shift in a medium can be
calculated in a much simpler way. To this end we evaluate the expectation
value of the current of the background fermions, $\langle\bar{\psi}_f
\gamma_{\mu} (C_V - C_A \gamma_5) \psi_{f`} \rangle$. The axial part
(the term proportional to $C_A$) vanishes if the medium is unpolarized so
that there is no preferred spin direction. The vector part is equivalent to
$(n_f-n_{\bar f})u_\mu$ where $n_f$ and $n_{\bar f}$ are the particle 
and antiparticle densities, respectively, and
$u$ is the medium's four--velocity.
Furthermore, we are only concerned with left--handed
neutrino fields for which $\gamma_5 \psi_{\nu}=-\psi_\nu$ or
$(1 - \gamma_5) \psi_{\nu}=2\psi_{\nu}$. In summary, the interaction 
Hamiltonian of Eq.~(\ref{eq:hamiltonian}) amounts to
\be
\sqrt{2} G_F C_V (n_f-n_{\bar f}) u_\mu\,\bar\psi_\nu\gamma^\mu\psi_\nu.
\ee
In the rest frame of the medium we have no preferred direction (no bulk
flows) so that $u=(1,0,0,0)$. Therefore, a left--handed neutrino in a
background medium feels a weak--interaction potential
\be
V = \pm \sqrt{2}G_F C_V' (n_f - n_{\overline{f}}), 
\ee
where the lower sign refers to anti--neutrinos. The dispersion relation is
\be
\omega=V+k,
\ee
which evidently does not resemble the one for a massive particle. Therefore,
one can not define an ``effective neutrino mass'' in the medium. 

The relevant coupling constants $C_V'$ for various background particles are
given in Table~\ref{kopplung}. For most cases, $C_V'$ is identical with the
neutral--current coupling $C_V$. However, if $f$ is the charged lepton
belonging to the neutrino, an additional term with $C_V = 1$ from the
Fierz--transformed charged current occurs.  For $f = \nu$ we have a factor 2
for the exchange amplitude of two identical particles.

\begin{table}[ht]
\begin{center}
\caption{\label{kopplung}Effective coupling constants for refraction of
neutrinos in a medium of background fermions. 
Note that $\sin^2 \Theta_W = 0.226 $.}
\medskip
\begin{tabular}{lll} \hline \hline
\noalign{\vskip2pt}
Fermion $f$ & Neutrino & $C_V'$\\ 
\noalign{\vskip2pt}
\hline    
\noalign{\vskip2pt}
Electron & $\en $ & $+\frac{1}{2} + 2 \sin^2 \Theta_W  $ \\[4pt] 
  & $\mun$, $\taun$ & $- \frac{1}{2} + 2 \sin^2 \Theta_W  $ \\[4pt] 
Proton & $ \en , \, \mun , \, \taun $ &  
$+\frac{1}{2} - 2 \sin^2 \Theta_W  $ \\[4pt] 
Neutron & $ \en , \, \mun , \, \taun $ &  
$ - \frac{1}{2} $ \\[4pt] 
Neutrino $\nu_a$& $ \nua $ &  
$ + 1  $ \\[4pt] 
& $ \nu_{b \ne a} $ &  
$ +\frac{1}{2}  $ \\[4pt] \hline
\end{tabular}
\end{center}
\end{table}

In an electrically neutral, normal
medium we have as many protons as electrons, at
least if the temperature is low enough that muons and pions are not present,
so that 
\be\label{schPot}
V = \pm\sqrt{2}G_F\times \cases{ 
-\frac{1}{2} n_n + n_e  &for $\en$,\cr
-\frac{1}{2} n_n        &for $\mun$ and $\taun$.\cr} 
\ee
The plus sign is for neutrinos, the minus sign for antineutrinos. For $n_e <
\frac{1}{2} \, n_n$ (e.g.~in a neutron star) the potential produced by the
medium is negative for neutrinos and positive for antineutrinos.  It is
important to note that the extra contribution for the electron flavor stems
from the charged--current interaction in $\en+e \ra \en + e$, which is not
possible for the other flavors.

\subsection{Exercises}

\subsubsection {Constraints on Neutrino Dipole Moments} 

\label{aufg1}

Neutrinos may have anomalous electric and magnetic dipole and transition
moments, which are small in the standard model but can be large in certain
extensions so that they have to be constrained.  The Lagrangian is
\be
\mbox{$\cal L$}_{\rm int} = \frac{1}{2} 
\sum\limits_{a,b} \left( \mu_{ab} \bar{\psi}_a \sigma_{\mu \nu} 
\psi_b + \epsilon_{ab} \bar{\psi}_a \sigma_{\mu \nu} \gamma_5 
\psi_b \right) F^{\mu \nu} , 
\ee
where the indices $a$ and $b$ run over the neutrino families, $\mu_{ab}$ is a
magnetic, $\epsilon_{ab}$ an electric transition moment, respectively, and a
static magnetic or electric dipole moment for $a = b$.  (Note that electric
dipole moments are CP violating.)  These moments are measured in units of the
Bohr magneton $\mu_B = e/2m_e$. $F^{\mu \nu}$ is the electromagnetic field
tensor and $\sigma_{\mu \nu} = \frac{i}{2}(\gamma_{\mu} \gamma_{\nu} -
\gamma_{\nu} \gamma_{\mu})$.

\begin{itemize}
  
\item[a)] Calculate the decay width $\gamma \to\nu\bar\nu$ for a neutrino
  family with a magnetic dipole moment $\mu$, when photons have an effective
  mass $\omega_{\rm p}$, as seen in Eq.~(\ref{gamdisrel}).
  
\item[b)] Calculate the energy loss rate in neutrinos of a nonrelativistic
  plasma at the temperature $T$.
  
\item[c)] The cores of low--mass red giant stars (about $0.5~M_\odot$, solar
  mass $M_{\odot} = 2 \times 10^{33}$~g) have an average density of
  approximately $2 \times 10^5~{\rm g~cm}^{-3}$ and are almost isothermal at
  $10^8~\rm K$. In order not to delay helium ignition too much, the neutrino
  loss rate $\epsilon$ is not allowed to exceed about
  $10\rm~erg~g^{-1}~s^{-1}$.  Which limit is obtained for $\mu$?
  
\item[d)] This limit is also valid for transition moments, which can lead to
  decays like $\nu_2 \ra \nu_1 + \gamma$. Why does the direct search for these
  radiative decays of massive neutrinos make no sense, provided one believes
  in the bound for $\mu$ obtained above?
  
\item[e)] Estimate or calculate a similar bound for a hypothetical electric
  charge (millicharge) of a neutrino.

\end{itemize}

\subsubsection*{Hints}

Work in the rest frame of the medium. Show that the squared and spin
summed matrix element is of the form
\be 
\left| \mbox {$\cal M$}\right|^2 
= M_{\alpha \beta} P^{\alpha} {\bar P}^{\beta}, 
\ee 
where $P$ and $\bar P$ are the four--momenta of the neutrino and
antineutrino, respectively, and 
\be 
M_{\alpha \beta} = 4 \mu^2 \left(
2 K_{\alpha} K_{\beta} - 2 K^2 \epsilon_{\alpha}^{\ast}
\epsilon_{\beta} - K^2 g_{\alpha \beta} \right) , 
\ee 
where $K=(\omega,{\bf k})$ is the photon four momentum and $\epsilon$
its polarization four vector.  We have $\epsilon_{\alpha}^{\ast}
\epsilon^{\alpha} = - 1 $ and $\epsilon_{\alpha} K^{\alpha} = 0$.  For
the neutrino phase--space integration use Lenard's formula 
\be
\int\frac{d^3{\bf p}}{2E_{\bf p}}\,\frac{d^3{\bf \bar p}}{2E_{\bf \bar
p}}\, P^\alpha{\bar P}^\beta\,\delta(K-P-\bar P) =\frac{\pi}{24}\,(K^2
g^{\alpha\beta}+2K^\alpha K^\beta).  
\ee 
The decay width is finally found to be 
\be 
\Gamma_{\gamma \ra \nu
\bar{\nu}} = \frac{\mu^2}{24 \pi} \frac{(\omega^2-k^2)^2}{\omega}.
\ee 
Note that the decay would be impossible in vacuum where $\omega=k$.
In the present situation, however, we may insert the dispersion
relation $\omega^2-k^2=\omega_{\rm p}^2$ to obtain the decay rate.

Since every photon decay liberates the energy $\omega$ in the form of 
neutrinos, the energy--loss rate per unit volume is  
\be
Q = g \int \frac{d^3 k}{(2 \pi)^3}\, f_k\, \omega\, 
\Gamma_{\gamma\to\nu\bar\nu} , 
\ee
with the photon distribution function $f_k$ (Bose--Einstein) and $g=2$ the
number of polarization states.  Useful integrals for this exercise are given
in Table~\ref{integrale}.

The matrix element and the width for the radiative neutrino decay $\nu_2 \to
\nu_1 + \gamma$ (transition dipole moment $\mu_{12}$) are
\be
\ba
\left| \mbox {$\cal M$} \right|^2 = 8 \mu_{12}^2 
(K \cdot P_1) (K \cdot P_2) , \\ \\
\Gamma_{\nu_2 \ra \nu_1 + \gamma} = \frac{\D \mu_{12}^2}{\D 8 \pi} 
\left( \frac{\D m_2^2 - m_1^2}{\D m_2} \right)^3 ,  
\ea
\ee
where $P_i$ denotes the momentum of neutrino $\nu_i$ with mass $m_i$ and 
$K$ the photon momentum.

\subsubsection{Supernova Neutrinos and Refraction}

A type II supernova explosion is actually the implosion of the burnt--out 
iron core (mass around 1.4~$M_{\odot}$) of a massive star. 
This collapse leads to a compact object with nuclear density 
($3 \times 10^{14}~\rm g~cm^{-3}$) and a radius of about 12~km.

\begin{itemize}

\item[a)] What is the gravitational binding energy?
  
\item[b)] 99\% of this energy is emitted in $\nu$'s and $\bar\nu$'s of all
  flavors. When the time for this process is 10~s, what is the luminosity (in
  erg s$^{-1}$) in one neutrino degree of freedom?

\item[c)] The average energy of the emitted neutrinos is  
\be \label{avE}
\langle E_{\nu} \rangle = \cases{ 
10~{\rm MeV} & for $\en$, \cr
14~{\rm MeV} & for $\bar\nu_e$, \cr
20~{\rm MeV} & other. \cr}
\ee
What is the number flux at the time of emission? What is the local 
neutrino density (per flavor) as a function of the radius above the neutron 
star surface?

\item[d)] At the surface of the neutron star the matter density falls off
  steeply. Assume that is follows a power law $\rho = \rho_R (R/r)^p$ with $p
  = 3-7$ and $\rho_R = 10^{14}~\rm g~cm^{-3}$. How does the electron
  density compare with the neutrino density during their emission?  Assume that
  the medium has as many protons as neutrons.
  
\item[e)] Compare the weak potential produced by the neutrinos with the one
  produced by normal matter. Assume that the energy flux is the same in all
  flavors, but the average energy is not, see Eq.\ (\ref{avE}). 
  Therefore, only for the
  electron flavor a difference between particles and antiparticles exists and
  thus a net contribution to the weak potential. Another important point is
  that the $\nu$'s are emitted almost collinear so that the background
  medium of neutrinos is not isotropic relative to a test neutrino; for an
  exactly pointlike source there would be no contribution at all.

\end{itemize}

\subsubsection{Neutrino Refraction in the Early Universe}

The ``normal'' neutrino refractive index is calculated on the basis of the
Fermi interaction (current--current interaction).  It can be interpreted as a
weak potential for the neutrinos. The medium in the early universe is almost
CP--invariant ---all particles have almost the same number density as their
antiparticles. Thus the refractive index nearly cancels to this order.  A weak
potential arises only from the matter--antimatter asymmetry of $\eta \simeq 3
\times 10^{-10}$ baryons per photon.

\begin{itemize}
  
\item[a)] Using this value for $\eta$, estimate the ``normal'' refractive
  index of neutrinos in the radiation dominated era before $e^{+}e^{-}$
  annihilation.  Use dimensional arguments to express $n_{\rm refr}$ as a
  function of the cosmic temperature $T$.  (Hint: the number density of
  relativistic thermal particles is proportional to $T^{3}$.)

\item[b)] Which Feynman diagrams contribute to neutrino forward 
scattering and thus to the refractive index?

\item[c)] The gauge boson propagator can be expanded if the 
momentum transfer $Q$ is small relative to the gauge boson mass, 
\be
D_{\mu \nu} = \frac{g_{\mu \nu}}{M_{Z,W}^2} + 
\frac{Q^2 g_{\mu \nu} - Q_{\mu} Q_{\nu}}{M_{Z,W}^4} + \ldots . 
\ee
The first term provides the Fermi theory of weak interactions. For which
diagram is the current--current term the exact result? For which diagrams does
one have to take higher terms into account?

\item[d)] Can you imagine other corrections which might be as important as the
  propagator expansion?
  
\item[e)] Estimate, again in form of a dimensional analysis, the contribution
  of the higher terms in the early universe. Compare with a). Interpretation?

\end{itemize}

\subsubsection*{Remark} 

If the medium consists of neutrinos of all flavors and of $e^+ e^-$, an exact
calculation for the CP asymmetric contribution in the early universe gives
$n_{\rm refr} - 1 = \xi G_F^2 T^4 /\alpha$ with $\xi = \frac{14}{45} \pi (3 -
\sin^2 \Theta_W) \sin^2 \Theta_W \simeq 0.61$ for $\en$ or $\bar\nu_e$, while
for the other flavors one finds $\xi = \frac{14}{45} \pi (1 - \sin^2 \theta_W)
\sin^2 \Theta_W \simeq 0.17$.

\subsubsection{The Sun as a Neutrino Lens}

The neutrino refractive index in media is important for oscillation phenomena. 
Can it be responsible for conventional refractive effects? 
Estimate the deflection angle of a neutrino beam when it hits 
a spherical body with a given impact parameter. Give a crude numerical 
value for the Sun. Compare with gravitational deflection.

\subsubsection*{Hints}

Assume parallel layers of the medium and a beam which propagates at an angle
$\alpha$ relative to the density gradient.  The refraction law informs us that
$n_{\rm refr} \sin \alpha =$ const, where $n_{\rm refr}$ is the refractive
index. Differentiating and some manipulations give for the beam deflection
\be
 \frac{d \alpha}{ds} = \frac{ \left| \nabla_{\perp} n \right|}{n_{\rm refr}}, 
\ee
where $s$ is a coordinate along the beam and $\nabla_{\perp}$ the gradient
transverse to the local beam direction. Since $| n_{\rm refr} - 1| \ll 1$ one
can take $n_{\rm refr} = 1$ in the denominator. The deflection is so small
that to lowest order the beam travels on a straight line.

The neutrino refractive index is $n_{\rm refr} = 1 - m_{\nu}^2/2E_{\nu}^2 -
V/E$, where $V$ is the weak potential of Eq.~(\ref{schPot}). Numerically one
finds $\sqrt{2}G_F\rho/m_u = 0.762 \times 10^{-13} \rm~eV \, \rho
/(g~cm^{-3})$ with the mass density $\rho$ and the atomic mass unit $m_u$.
The density at the center of the Sun is about 150 $\rm g~cm^{-3}$ and the
radius of the Sun is $R_{\odot} = 6.96 \times 10^{10}$~cm.

Gravitation affects the beam deflection through the ``refractive index''
$n_{\rm refr} = 1 - 2 \Phi$, where $\Phi$ is the gravitational potential. In
natural units, Newton's constant is $G_N = 1/m_{\rm Pl}^2$ with the Planck
mass $m_{\rm Pl} = 1.22 \times 10^{19}$ GeV.

\newpage


\section{Neutrino Oscillations}

\subsection{Vacuum Oscillations}

If neutrinos have nonzero masses --- and thus have properties beyond the
standard model --- they can oscillate between the flavor eigenstates.  A flavor
eigenstate is operationally defined as a neutrino state which appears in
association with a given charged lepton. For example, the anti--neutrino
emerging from neutron decay $n\ra p+e^-+\bar\nu_e$ is by definition an
electron anti--neutrino. Likewise, the reaction $\mu^-+p\ra n+\nu_\mu$ produces
a muon neutrino. Within the standard model, flavor--changing neutral currents
do not exist, i.e.\ in a scattering process of the form $\nu+n\ra n+\nu$ the
out--going neutrino has exactly the same flavor as the incoming one.  However,
in analogy to the quark sector, the flavor eigenstates need not be eigenstates
of the mass operator.  If neutrinos have masses at all, it is generally
assumed that the mass operator violates the conservation of individual
lepton--flavor numbers.

The simplest example is that of two lepton families.  The mass eigenstates
$\nu_j$, $j = 1$, 2 are connected to the flavor eigenstates $\nual$ and
$\nube$ via
\be \label{mm}
\left( \ba \nual \\ \nube \ea \right) = 
\left( \baz \cos \theta & \sin \theta \\ 
            -\sin \theta & \cos \theta \ea \right) 
\left( \ba \nu_1 \\ \nu_2 \ea \right) . 
\ee
Depending on the context, $\nu_j$ may stand for the neutrino field operator or
simply for a neutrino state, in which case $\nu_j$ stands for $|\nu_j\rangle$.
The mixing matrix is unitary and therefore has one nontrivial free parameter,
the mixing angle $\theta$. For three ($n$) 
families, the mixing matrix would have
three [$\frac{1}{2} n (n - 1)$]  
nontrivial angles $\theta_j$. In addition, for Dirac neutrinos it has
one [$\frac{1}{2} (n - 2) (n - 1)$] 
CP--violating phase(s), and three [$\frac{1}{2} n (n - 1)$] nontrivial 
phases for Majorana neutrinos.
However, it can be shown that in oscillation experiments only the 
number of phases given by the Dirac case, i.e.~one 
[$\frac{1}{2} (n - 2) (n - 1)$] can be 
measured. 
We have therefore the same structure as for the Cabbibo--Kobayashi--Maskawa
(CKM) matrix in the quark sector of the standard model.

\begin{figure}[t]
\setlength{\unitlength}{1cm}
\begin{center}
\epsfig{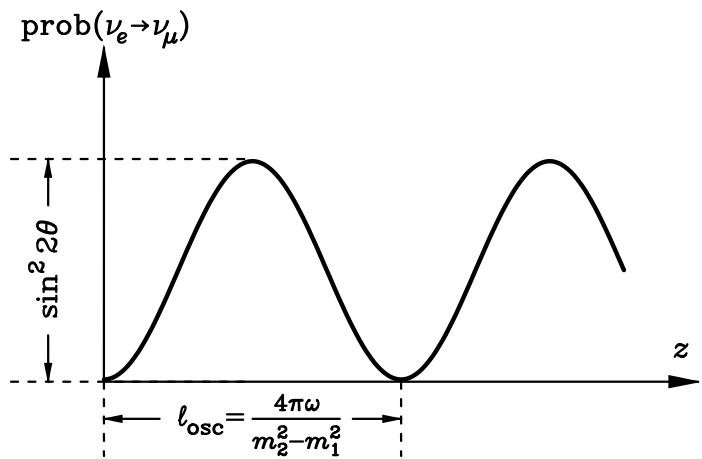}
\end{center}
\caption{\label{oscillation}Oscillation pattern for two--flavor oscillations
(neutrino energy $\omega$, distance $z$).}
\end{figure}

It is easy to see how neutrino oscillations arise if we imagine that a
neutrino with energy $E$ of a given flavor, for example $\nu_e$, is produced
at some location ${\bf x}=0$. It can be decomposed into mass eigenstates
according to Eq.~(\ref{mm}) so that
\be\label{eq:statezero}
\nu(0)=\nu_e=\cos\theta\,\nu_1+\sin\theta\,\nu_2.
\ee
We imagine the mass eigenstates to propagate as plane waves so that each
of them evolves along the beam as
\be
\nu_j\,e^{-i(Et-{\bf p}_j\cdot {\bf x})}
\ee
with $j=1$ or 2. Here,  
\be
p_j=|{\bf p}_j|=(E^2-m_j^2)^{1/2}\simeq E-\frac{m_j^2}{2E}\,,
\ee
where the last approximation holds in the relativistic limit 
$m_j\ll E$. This is surely justified since one expects $m_j$ 
to be smaller than a few eV and typical energies start at a few MeV\@. 
Therefore, the original state of 
Eq.~(\ref{eq:statezero}) evolves as
\be\label{eq:evolution}
\nu(x)=e^{-i E(t-x)}
\left(\cos\theta\,e^{i(m_1^2/2E)x}\,\nu_1+\sin\theta\,
e^{i(m_2^2/2E)x}\,\nu_2\right),
\ee
where $x=|{\bf x}|$. Next, we can invert Eq.~(\ref{mm}) to express
the mass eigenstates in terms of the flavor eigenstates, and insert these
expressions into Eq.~(\ref{eq:evolution}). Assuming the other flavor is
$\nu_\mu$, one then finds immediately that
the $\nu_\mu$ amplitude of the beam evolves as
\be 
\langle \nu_{\mu} | \nu_e \rangle = 
\frac{1}{2} \sin 2\theta\, \left( e^{i (m_2^2/2E)\,x}
 -  e^{i (m_1^2/2E)\,x} \right)\, e^{-i E(t-x)}. 
\ee
Squaring this amplitude gives us the probability for observing a $\nu_\mu$
at a distance $x$ from the production site  
\be \label{prob}
P(\nu_e \ra \nu_\mu) = \sin^2 2 \theta \,\, 
\sin^2\left( \frac{x\, \delta m^2}{4 E}\right), 
\ee
where $\delta m^2\equiv m_2^2-m_1^2$.  
A mono--energetic neutrino beam
thus oscillates with amplitude $\sin^2 2\theta$ and wave number
$k_{\rm osc} = \delta m^2 /4E$ (Fig.~\ref{oscillation}).  The maximum
effect occurs for $\theta = \pi/4$.  One usually defines the
oscillation length
\be
\ell_{\rm osc}=\frac{4\pi E}{\delta m^2}=2.48~{\rm m}
\,\frac{E}{1~{\rm MeV}}\,\frac{1~{\rm eV}^2}{\delta m^2}.
\ee
The neutrino beam has returned to its original state after traveling a
distance $\ell_{\rm osc}$. The probability for finding the neutrino in its
original state after traveling a distance $x$ is $P(\nu_e \ra \nu_e) = 1 - P(
\nu_e \ra \nu_\mu)$.

Note that oscillations would be impossible for completely degenerate masses
($m_1^2 = m_2^2$), including the case of vanishing neutrino masses, or for a
vanishing mixing angle.

The generalization of these results to three or more families is
straightforward but complicated; it can be found in many textbooks,
e.g.\ in Ref.~\cite{schmitz}. Equation~(\ref{prob}) then reads 
\be
 P( \alpha \ra \beta) = \delta_{\alpha\beta} - 2 \, \Re \sum_{j>i} 
U_{\alpha i}U_{\alpha j}^{\ast} U_{\beta i}^{\ast} U_{\beta j} 
\left[ 1 - \exp\left(-i \frac{\Delta_{ij}}{4 E}\,x\right) \right]
\ee
with $\Delta_{ij}=m_i^2 - m_j^2$.  In general $P( \alpha \ra \beta)
\neq P( \beta \ra \alpha)$ due to the complex phase of the unitary
mixing matrix $U$, offering a possibility to measure this phase. A
convenient parametrisation of $U$ is
\be
U =  \left( 
\bad c_1 & s_1 c_3 & s_1 s_3 \\[0.2cm]
     -s_1 c_2 & c_1 c_2 c_3 - s_2 s_3 e^{i \delta} & c_1 c_2 s_3 + s_2 c_3 
e^{i \delta} \\[0.2cm] 
s_1 s_2 & - c_1 s_2 c_3 - c_2 s_3 e^{i \delta} & -c_1 s_2 s_3 + c_2 c_3 
e^{i \delta} 
\ea
\right) 
\ee
with $s_i = \sin \theta_i \mbox{ and } c_i = \cos \theta_i$. For
Majorana neutrinos $U$ has to be multiplied with 
${\rm diag}(e^{i\lambda_1},e^{i\lambda_2},1)$.

Thus far we have assumed that the neutrinos are mono--energetic 
and the sources
and detectors are pointlike.  Since nature is not so kind as to provide us
with these simple cases, one has to convolute these formulas with energy and
distance distributions. Naturally, these effects smear out the signature of
oscillations, thereby complicating the interpretation of the experiments.  A
given experiment or observation usually provides exclusion or evidence regions
in the parameter plane of $\sin^2 2\theta$ and $\delta m^2$.

\subsection{Oscillation in Matter}

\subsubsection{Homogeneous Medium}

Neutrino oscillations arise over macroscopic distances because the momentum
difference $\delta m^2/2E$ between two neutrino mass eigenstates of energy $E$
is very small for neutrino masses in the eV range or below and for energies in
the MeV range or above. Wolfenstein was the first to recognize that the
neutrino refractive effect caused by the presence of a medium can cause a
momentum difference of the same general magnitude, implying that the extremely
small weak potential for neutrinos in a medium can modify neutrino
oscillations in observable ways.

In order to understand how the neutrino potential enters the oscillation
problem, it is useful to back up and derive a more formal equation for the
evolution of a neutrino beam. To this end we begin with the Klein--Gordon
equation for the neutrino fields
\be
(\partial_t^2-\nabla^2+M^2)\Psi=0
\ee
where in the general three--flavor case 
\be
M^2=\pmatrix{m_1^2&0&0\cr0&m_2^2&0\cr0&0&m_3^2\cr}
\hbox{\qquad and\qquad}
\Psi=\pmatrix{\nu_1\cr\nu_2\cr\nu_3\cr}.
\ee
If we imagine neutrinos to be produced with a fixed energy $E$ at some source,
their wave functions vary as $e^{-iEt}$ so that their spatial propagation is
governed by the equation
\be
(-E^2-\partial_x^2+M^2)\Psi=0,
\ee
where we have reduced the spatial variation to one dimension, i.e.\ we
consider plane waves. 

In the relativistic limit $E\gg m_j^2$ we may linearize this wave equation by
virtue of the decomposition
$(-E^2-\partial_x^2)=-(E+i\partial_x)(E-i\partial_x)$.  Since
$-i\partial_x\nu_j=p_j\nu_j$ with $p_j=(E^2-m_j^2)^{1/2}\simeq E$ it is
enough to keep the differential in the difference term, while replacing it
with $E$ in the sum, leading to $(-E^2-\partial_x^2)\ra -2E(E+i\partial_x)$.
Therefore, in the relativistic limit the evolution along the beam is governed
by a Schr\"odinger--type equation 
\be 
i\partial_x\Psi=(-E+\Omega)\Psi,\qquad \Omega=\frac{M^2}{2E}.  
\ee 
Usually this sort of equation is written down for the time--variation instead
of the spatial one so that it looks more like a conventional Schr\"odinger
equation.  However, in the problem at hand we ask about the variation of the
flavor content along a stationary beam, so that it is confusing to use a
differential equation for the time variation which then has to be
re--interpreted as describing the evolution along the beam. Either way,
the main feature of this equation is that it is a complex linear equation
involving a ``Hamiltonian'' matrix $\Omega$; the term $-E$ contributes a
global phase which is irrelevant for the oscillation probability. 

The potential caused by the medium is then easily included by adding it to the
Hamiltonian
\be
\Omega\ra\Omega_{\rm M}=\frac{M^2}{2E}+V
\ee
where $V$ is a matrix of potentials which is diagonal in the
weak--interaction basis with the entries given by Eq.~(\ref{schPot}).

As a specific example we now consider two--flavor mixing between $\nu_e$ and
$\nu_\mu$, and write down the Hamiltonian in the weak interaction basis.  It
is connected to the mass basis by virtue of the unitary transformation
$\nu_\alpha=U_{\alpha i}\nu_i$ of Eq.~(\ref{mm}).
The squared mass matrix then transforms as $UM^2U^\dagger$, leading in the
weak basis explicitly to
\be
M^2=\frac{\Sigma}{2}+\frac{\delta m^2}{2}
\pmatrix{-\cos2\theta&\sin2\theta\cr\sin2\theta&\cos2\theta\cr},
\ee
where $\Sigma=m_2^2+m_1^2$ and $\delta m^2=m_2^2-m_1^2$. Including the
weak potential then leads to the ``Hamiltonian''
\begin{eqnarray}
\Omega_{\rm M}&=&\frac{\Sigma}{4E}+\frac{\delta m^2}{4E}
\pmatrix{-\cos2\theta&\sin2\theta\cr\sin2\theta&\cos2\theta\cr}
+\sqrt{2}G_F\pmatrix{n_e-\frac{1}{2}n_n&0\cr0&-\frac{1}{2}n_n\cr}\nonumber\\
&=&
\frac{1}{2}\left[\frac{\Sigma}{2E}+\sqrt2\,G_F(n_e-n_n)\right]\nonumber\\
&+&\frac{1}{2}\left(\sqrt2\,G_Fn_e-\frac{\delta m^2}{2E}\cos 2 \theta\right)
\pmatrix{1&0\cr0&-1\cr}+
\frac{1}{2}\,\sin2\theta\,\pmatrix{0&1\cr1&0\cr}
\end{eqnarray}
which governs the oscillations in a medium. The first term which is
proportional to the unit matrix produces an irrelevant overall phase so that
the medium effect on the oscillations depends only on the electron density
$n_e$, i.e.\ on the {\it difference\/} between the weak potentials for
$\nu_e$ and $\nu_\mu$. Recall that this difference arises from the
charged--current piece in the $\nu_e$ interaction with the electrons of the
medium. 

The meaning of this complicated--looking expression becomes more transparent if
one determines the ``propagation eigenstates,'' i.e.\ the basis where
$\Omega_{\rm M}$ is diagonal. In vacuum we have $M^2=2E\Omega$ so that in the
medium one may define an effective mass matrix by virtue of $M^2_{\rm
  M}=2E\Omega_{\rm M}$.  The eigenvalues $m_{\rm M}^2$ of this matrix are
found in the usual way by solving ${\rm det}(2 E \Omega_{\rm M} - m_{\rm
  M}^2)=0$.  The two roots and their difference are found to be
\bea \label{dmm2} 
m_{\rm M}^2
= \frac{\D 1}{\D 2} \left( \Sigma + 2 \sqrt{2} G_F (n_e - n_n) E \mp
  \delta m_{\rm M}^2 \right) , \\[0.8cm]
\delta m_{\rm M}^2 = \left[ (\delta m^2)^2 + 4 E G_F n_e \left(2 E G_F n_e -
    \sqrt{2}\, \delta m^2 \cos 2 \theta \right) \right]^{1/2}.  
\eea 
In vacuum where $n_e = n_n = 0$ these expressions reduce to $m^2_{1,2}$ and
$\delta m^2=m_2^2-m_1^2$. We stress that the in--medium effective squared
``masses'' should not be literally interpreted as effective masses as they
depend on energy; $m^2_{\rm M}$ may even become negative. The ``effective
masses'' are simply a way to express the dispersion relation in the medium.

The transformation between the in--medium propagation eigenstates and the
weak--interaction eigenstates is effected by a unitary transformation of the
form Eq.~(\ref{mm}) with the in--medium mixing angle
\be \label{thetam}
\tan 2 \theta_{\rm M} =
\frac{\D \sin 2 \theta}
{\D  \cos 2 \theta - A } 
\ee
which is equivalent to
\be
\sin^2 2 \theta_{\rm M} = \frac{\D \sin^2 2 \theta}
{\D (\cos  2 \theta-A)^2 + \sin^2 2 \theta}.
\ee
Here
\be
A \equiv \frac{2 \sqrt{2}E G_F n_e}{\delta m^2} = 1.52 \times 10^{-7} 
\frac{Y_e \, \rho}{\rm g \, cm^{-3}}\, \frac{\rm E}{\rm MeV}\, 
\frac{\rm eV^2}{\delta m^2} ,
\ee
where $Y_e$ is the electron number per baryon and $\rho$ the mass density.

With these results it is trivial to transcribe the oscillation probability
from the previous section to this case,
\be
P_{\rm M} (\en  \ra \mun) = \sin^2 2 \theta_{\rm M} 
\sin^2\left( \frac{x\,\delta m_{\rm M}^2}{4 E}\right) . 
\ee
Evidently we have
\be \label{Lm}
\ell_{\rm M} = \frac{ 4\pi E }{\delta m_{\rm M}^2} 
= \frac{\sin 2 \theta_{\rm M} }{\sin 2 \theta}\,
\ell_{\rm vac}, 
\ee 
for the in--medium oscillation length. 

\subsubsection{\label{MSW}MSW Effect}

The mixing angle in matter, $\sin^2 2 \theta_{\rm M}$, is a function of $n_e$
and $E$. It becomes maximal ($\theta_{\rm M}=\pi/4$) when $A =A_R= \cos 2
\theta$.  Here $\delta m_{\rm M}^2$ has a minimum, the oscillation length
$\ell_{\rm M}$ a maximum. Even if the mixing angle in vacuum is small, 
on resonance the in--medium mixing is maximal,
{\it independently} of the mixing angle in vacuum.

The most interesting situation arises when a neutrino beam passes through a
medium with variable density, the main example being solar neutrinos which are
produced at high density in the solar core.  Considering the case $m_1^2
\simeq 0$ and $\theta$ small, we find for high density ($A\gg A_R$) that
$\theta_{\rm M}\simeq \frac{\D \pi}{\D 2}$, implying
\begin{eqnarray}
&&\nu_{1\rm M}  \simeq -\mun,\quad  m_{1\rm M}^2 \simeq 0,\nonumber\\
&&\nu_{2\rm M}  \simeq +\en ,\quad  m_{2\rm M}^2 \simeq A\, \delta m^2 . 
\end{eqnarray}
On the other hand, for low densities ($A\ll A_R$) 
we have vacuum mixing ($\theta_{\rm M}\simeq \theta\ll 1$)
so that 
\begin{eqnarray}
&&\nu_{1\rm M}  \simeq +\en,\quad  m_{1\rm M}^2 \simeq 0,\nonumber\\
&&\nu_{2\rm M}  \simeq +\mun ,\quad  m_{2\rm M}^2 \simeq m_2^2. 
\end{eqnarray}
Therefore, the propagation eigenstate $\nu_{2\rm M}$  
which at high density is approximately
a $\nu_e$ turns into a $\nu_\mu$ at low density, and vice versa. Therefore,
if the neutrino is born as a $\nu_e$, and if the density variation is 
adiabatic so that the neutrino can be thought of being in a propagation
eigenstate all along its trajectory, it will emerge as a $\nu_\mu$.

In order for this resonant conversion to occur, two conditions must
be met. First, the production and detection must occur on opposite sides of
a layer with the resonant density. In the Sun, the neutrinos are produced
at high density so that we need to require
\be
A (\mbox{place of production}) > A_R = \cos 2 \theta . 
\ee
This can be rewritten as a constraint on the neutrino energy, 
\be \label{mswenergy}
E > E_0 = \frac{\delta m^2 \cos 2 \theta}{2 \sqrt{2} G_F n_e} 
= 6.6 \times 10^6~{\rm MeV}\,\cos 2 \theta\, \frac{\delta m^2}{\rm eV^2} 
\frac{\rm g~cm^{-3}}{Y_e \rho}. 
\ee
Second, for the neutrino to stay in the state $\nu_{2m}$, 
the density gradient has to be moderate, i.e.\ the density variation
must be small for several oscillation lengths. This condition can
be expressed as $\ell_{\rm M}^{-1}\gg\nabla\ln n_e$. One often
defines the adiabaticity parameter
\be \label{mswadia}
\gamma = \frac{\D \delta m^2}{\D 2 E} \frac{\D \sin^2 2 \theta}
{\D \cos 2 \theta} \frac{\D 1}
{\D |\nabla \ln n_e|}
\ee
so that adiabaticity is achieved for $\gamma\ll1$. This condition must be
met along the entire trajectory. As the oscillation length is longest on
resonance when the mixing is maximum, the adiabaticity condition is
most restrictive on resonance. 
In general one finds a triangle in $ \sin^2 2 \theta$--$\delta m^2$ space
where these conditions are fulfilled ---see Exercise \ref{mswaufgabe}.
In Fig.~\ref{mixing} the triangular contours show the range of masses
and mixing angles where the solar $\nu_e$ flux is reduced to the measured
levels for experiments with different spectral response, i.e.\ which 
measure different average neutrino energies. 

\subsection{Exercises}

\subsubsection{\label{oscilauf}Polarization vector and neutrino oscillations}

Neutrino oscillations are frequently described by a Schrödinger equation 
of the form  
\be
i \dot{\Psi} = \Omega \Psi \mbox{\qquad with\qquad} 
\Omega = p + \frac{M^2}{2 p} , 
\ee
with $p$ the neutrino momentum, $M$ the mass matrix, and $\Psi$ a column
vector with two or more flavors. For two generations, the relation between
flavor and mass eigenstates is given by Eq.~(\ref{mm}).
Instead of the state vectors, however, one can work with the 
$2{\times}2$ density matrix in flavor space which is defined by 
\be
\rho_{ab} = \Psi^{\ast}_b \Psi_a , 
\ee
where the indices $a$ and $b$ run, for example, over $\en$ and $\mun$ or over
1 and 2 in the mass basis.  With the help of the density matrix we can find an
intuitive geometric interpretation of oscillation phenomena. In addition, one
can treat statistical mixtures of states, i.e.\ when the neutrinos are not
characterized by pure states.

\begin{itemize}
  
\item[a)] Show that the equation of motion is: $i \dot{\rho} = [\Omega, \rho]
  =[M^2, \rho]/2p $.
  
\item[b)] Write the mass matrix in the form $M^2/2p = V_0 - \frac{1}{2}
  {\bf V \cdot \sigma}$ and show, that in the flavor basis
\be
V_0 = \frac{m_2^2 + m_1^2}{4 p} \mbox{ and } 
{\bf V} = \frac{2 \pi}{\omega_{\rm osc}} \left( 
\ba \sin 2 \theta \\[0.2cm] 0 \\[0.2cm] \cos 2 \theta \ea \right) 
\mbox{ with } \omega_{\rm osc} = \frac{4 \pi p}{m_2^2 - m_1^2}.  
\ee
The vector ${\bf V}$ is thus rotated against the 3--axis with 
the angle $2 \theta$. Has this orientation in the 1--2 plain a physical 
meaning?

\item[c)] Express the density matrix in terms of a polarization vector in form
  of $\rho = \frac{1}{2}(1 + {\bf P} \cdot {\bf \sigma})$.  
Physical interpretation
  of its components?

\item[d)] Which property of ${\bf P}$ characterizes the ``purity'' of the
  state, i.e.\ when does the density matrix describe pure states, when
  maximally incoherent mixing?

\item[e)] Show that the equation of motion is a precession formula,
  $\dot{{\bf P}} = {\bf V} \cdot {\bf P}$. Obtain the oscillation probability
  for an initial $\en$.

\item[f)] The energy of (non--mixed) relativistic neutrinos in a normal medium
  is $E = p + (m^2/2p) + V_{\rm med}$. Here $V_{\rm med}$ is given by
  Eq.~(\ref{schPot}). What is ${\bf P}$ in the medium?  What is the mixing
  angle in the medium?
  
\item[g)] In a medium consisting of neutrinos (supernova, early universe) one
  can not distinguish between a test neutrino and a background neutrino, so
  that oscillations with medium effects are in general nonlinear.  What is the
  advantage of the density matrix formalism in this situation?

\end{itemize}

\subsubsection{MSW Effect in the Sun\label{mswaufgabe}}

The conditions for the MSW effect are given by the Eqs.~(\ref{mswenergy}) and
(\ref{mswadia}).  Determine the region in $ \sin^2 2 \theta$--$\delta m^2$
space, where one expects almost complete flavor inversion, i.e.\ the MSW
triangle. For this purpose, assume that all solar neutrinos are produced with
an energy of $E=1$~MeV and that the density profile of the Sun is approximated
by an exponential of the form $n_e = n_c \exp(-r/R_0)$, with a scale height of
$R_0 = R_{\odot}/10.54$ and a density at the center of $n_c = 1.6 \times
10^{26} \rm~cm^{-3}$.
 
\begin{figure}
\setlength{\unitlength}{1cm}
\begin{center}
\epsfig{file=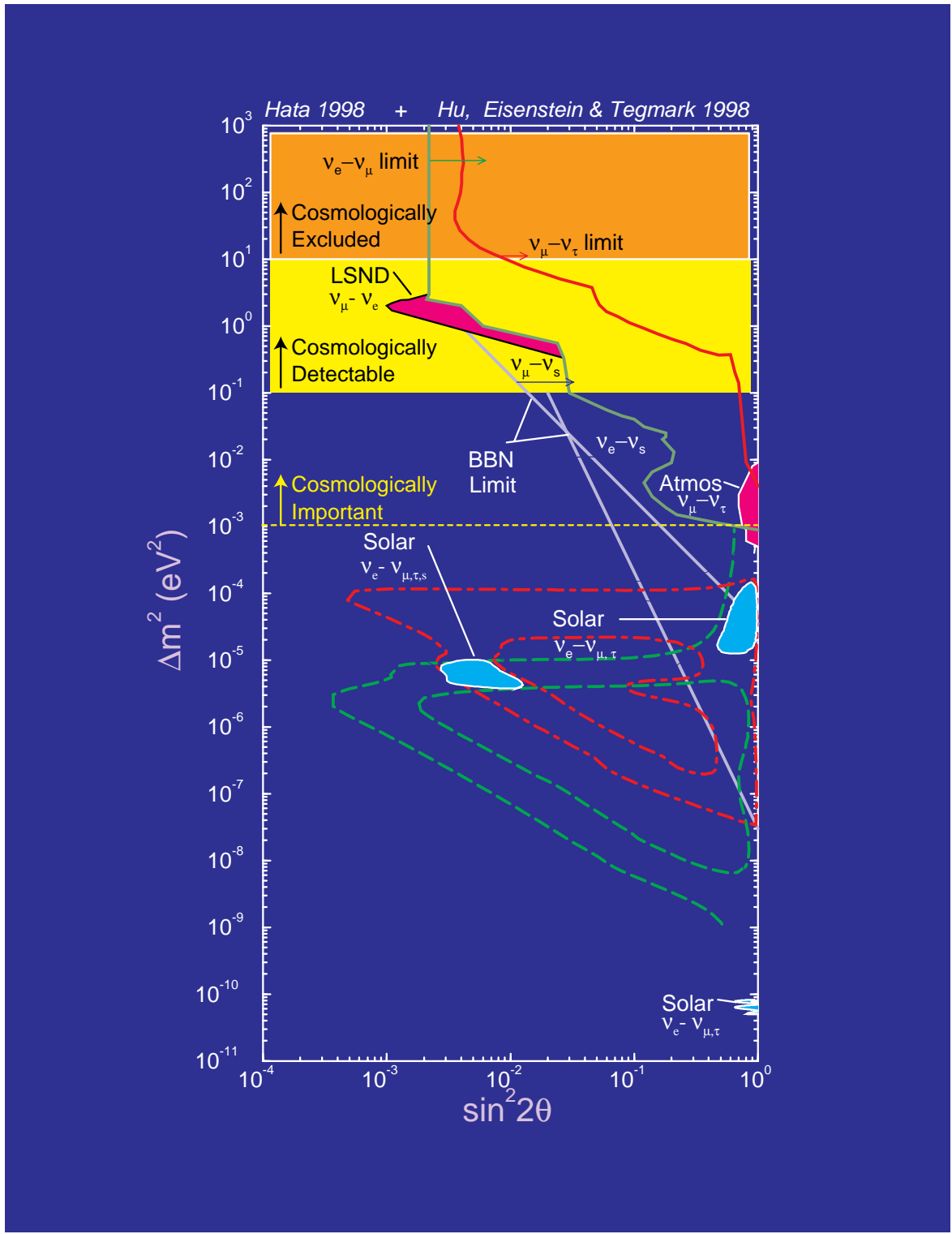,width=15.5cm}
\caption{\label{mixing} Limits and evidence for neutrino oscillations
(Figure courtesy of Max Tegmark).}
\end{center}
\end{figure}


\section{Experimental Oscillation Searches}

\subsection{Typical Scales}

We now turn to a discussion of some experimental strategies for the detection
of neutrino oscillations.  The most widely used neutrino sources are the Sun,
the Earth's atmosphere where neutrinos emerge from cosmic--ray interactions, or
man--made devices such as reactors and accelerators. One distinguishes between
appearance and disappearance experiments.  In the former, one searches for the
appearance of another flavor than has been produced in the source, while the
latter are only sensitive to a deficit of the original flux.

From Eq.~(\ref{prob}) one finds that an experiment with typical neutrino
energies $E$ and a distance $L$ between source and detector is sensitive to a
minimal value of the mass--squared difference of
\be
(\delta m^2)_{\rm min} \simeq \frac{E}{L}.  
\ee
Therefore, different experiments probe different regions of the mass
sector. In Table~\ref{Eoverx} we give some examples.
 
\begin{table}[ht]
\begin{center}
\caption{\label{Eoverx}Characteristics of typical oscillation experiments.}
\medskip
\begin{tabular}{lllll} \hline \hline
\noalign{\vskip4pt}
Source & Flavor & $E$ [GeV] & $L$ [km] & $(\delta m^2)_{\rm min}$ 
[eV$^2$]\\
\noalign{\vskip4pt}
 \hline
\noalign{\vskip4pt}
Atmosphere & $\stackrel{(-)}{\nu_e} , \; \stackrel{(-)}{\nu_\mu}$ 
& $10^{-1} \ldots 10^{2}$ & 
$10 \ldots 10^4$ & $10^{-6}$ \\
\noalign{\vskip4pt}
Sun & $\nu_e$ & $10^{-3} \ldots 10^{-2}$ & $10^{8}$ & $10^{-11}$ \\
\noalign{\vskip4pt} 
Reactor & $\bar\nu_e$ & $10^{-4} \ldots 10^{-2}$ & 
$10^{-1}$ & $10^{-3}$ \\
\noalign{\vskip4pt}
Accelerator & $\nu_e , \, \stackrel{(-)}{\nu_\mu} $ & $10^{-1} \ldots 1$ 
& $10^{2}$ & $10^{-1}$ \\
\noalign{\vskip4pt} 
\hline 
\end{tabular}
\end{center}
\end{table}

\subsection{Atmospheric neutrino experiments}

When cosmic rays, i.e.~protons and heavier nuclei interact with the Earth's
atmosphere they produce kaons and pions, which in turn decay into muons,
electrons and neutrinos.  Since the initial state is positively charged one
has more neutrinos than anti--neutrinos, but the experiments are insensitive
to this effect. On the other hand, the flavor can be very well measured; from
the simple production mechanism one expects
\be
N(\mun) :N(\en ) = 2:1 . 
\ee
This ratio depends on the energy of the measured neutrinos since the lifetime
of high-energy muons is increased by their Lorentz factor so that they may hit
the ground before decaying.  One often uses the double ratio
\be
R = \frac{\D \left( N(\mu)/N(e) \right)_{\rm meas}}
{\D \left( N(\mu)/N(e) \right)_{\rm MC}} , 
\ee
meaning the ratio of the measured flavor ratio divided by the expectation from
Monte--Carlo simulations.  The double ratio has the advantage that the
uncertainty of the overall absolute flux cancels out.  The average neutrino
energy is found to be $\langle E_{\nu } \rangle \simeq 0.6$ GeV.

\begin{figure}[b]
\setlength{\unitlength}{1cm}
\begin{center}
\epsfig{file=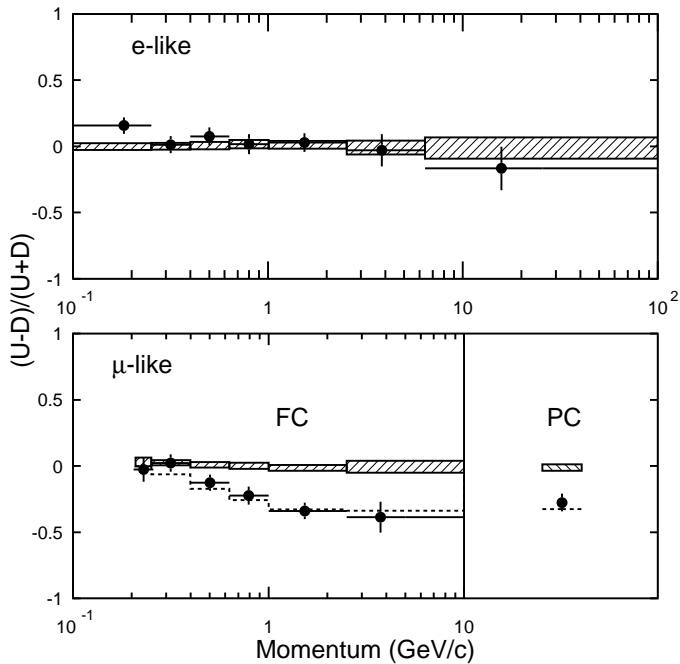,width=9.5cm}
\end{center}
\caption{\label{SKfig}Up--down asymmetry at SK. The hatched region 
is the expectation without oscillation, the dots the measurements, while 
the dashed line represents the best--fit oscillation case.
(Figure from \cite{SKdata}.)}
\end{figure}

A textbook example of an experiment of this kind is SuperKamiokande (SK)
\cite{SKdata}, an underground detector consisting of 50~kt of water,
surrounded by 11000 photomultipliers. Neutrinos react with the protons and
neutrons of the target and produce electrons and muons. These charged
particles are identified by their cones of Cherenkov light which are fuzzier
for the electrons.
Since cosmic rays are distributed almost isotropically and the atmosphere is
spherically symmetric, one expects the flux to be the same for down or
up--coming neutrinos.  However, it was found that up--coming muon neutrinos
are significantly suppressed.  This up--down asymmetry is shown in
Fig.~\ref{SKfig}. Plotted is the momentum of the charged leptons against
\be
A = \frac{U - D}{U + D}, 
\ee
where $U$ and $D$ are the number of upward and downward going events,
i.e.~their zenith angle is larger or smaller than about $78^\circ$,
respectively.  The dashed line corresponds to the best--fit mixing
parameters of SK, $\delta m^2 = 2.2 \times 10^{-3}$~eV$^2$ and $\sin^2 2
\theta = 1$. Roughly the same parameters are found by similar experiments,
like IMB \cite{IMB}, MACRO \cite{MACRO} or Soudan \cite{Soudan}. For the
double ratio $R$ values between $0.4$ and $0.7$ were found.

Therefore, the $\mun$'s probably oscillate into $\taun$'s. The other
possibility, oscillation into sterile neutrinos $\nu_s$ is disfavored because
the observed rate of the NC process $\nu N \to \nu \pi^0 X$ is about as
expected. Sterile neutrinos by definition do not take part in such
reactions. In addition, other properties such as the energy distributions of
the final--state charged leptons tend to confirm the $\nu_\mu$--$\nu_\tau$
interpretation.

\subsection{Accelerator Experiments}

Evidence for oscillations were present before the SuperKamiokande results.
The LSND \cite{LSND} collaboration used a 800 MeV proton beam colliding on a
water target so that pions were produced. The $\pi^-$ were captured in a
copper block while the $\pi^+$ decayed into $\mu^+ \mun$.  These in turn
decayed into a positron and two neutrinos.  Therefore, one expects the same
number of $\mun$, $\bar\nu_\mu$ and $\en$, but no $\bar\nu_e$.  In a
scintillation detector 30~m behind the source they looked for $\bar\nu_e$ in
the reaction $\bar\nu_e + p \to e^+ + n$. The experimental signature is thus
the Cherenkov cone from the positron and a photon from the reaction $\gamma +
p \to {\rm D} + \gamma \, (2.2 \rm \, MeV)$.

The LSND collaboration measured an excess of about 40 events above the
background.  The interpretation as oscillations, however, is controversial
because the very similar KARMEN \cite{KARMEN} experiment sees no events in
about the same parameter range. On the other hand, KARMEN will not be able to
exclude the LSND results.  The remaining parameter range where a consistent
interpretation as oscillations remains possible is shown in
Fig.~\ref{mixing}, i.e.\ $\delta m^2 \simeq 1 $ eV$^2$ and $\sin^2 2
\theta \simeq 10^{-2}$.

\subsection{Reactor Experiments}

In power reactors, nuclear fission produces $\bar\nu_e$ with energies of
typically several MeV\@. These energies are too low to produce a charged mu or
tau lepton in the detector so that reactor experiments are always
disappearance experiments.  The detection reaction is
$\bar\nu_e  + p \to e^+ + n$.

\begin{figure}[t]
\setlength{\unitlength}{1cm}
\begin{center}
\vspace{4cm}
\epsfig{file=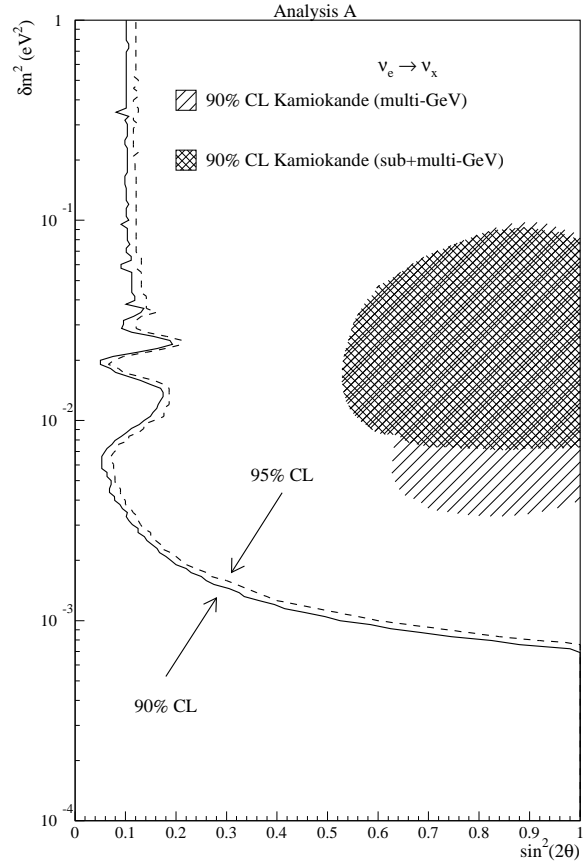,width=8.1cm}
\end{center}
\caption{\label{choozfig}Exclusion plot of the CHOOZ experiment; 
the area to the right of the lines is excluded.  
Also shown is the allowed parameter space of Kamiokande.
The SK has shifted these values 
towards lower masses, yet they still lie inside the forbidden area. 
(Figure from \cite{CHOOZ}.)}
\end{figure}

Thus far, none of the reactor experiment gives evidence for oscillations.
However, they have produced very important exclusion areas in the $\sin^2 2
\theta$--$\delta m^2$ space.  Most importantly, the CHOOZ
experiment~\cite{CHOOZ} has excluded a large range of mixing parameters 
(Fig.~\ref{choozfig})
so that a putative $\nu_\mu$--$\nu_e$ oscillation interpration of the
SuperKamiokande results is inconsistent with their limits. 

\subsection{\label{sunexp}Solar Neutrino Experiments}

\begin{figure}[b]
\setlength{\unitlength}{1cm}
\hbox to\hsize{\hss
\hbox{\epsfig{file=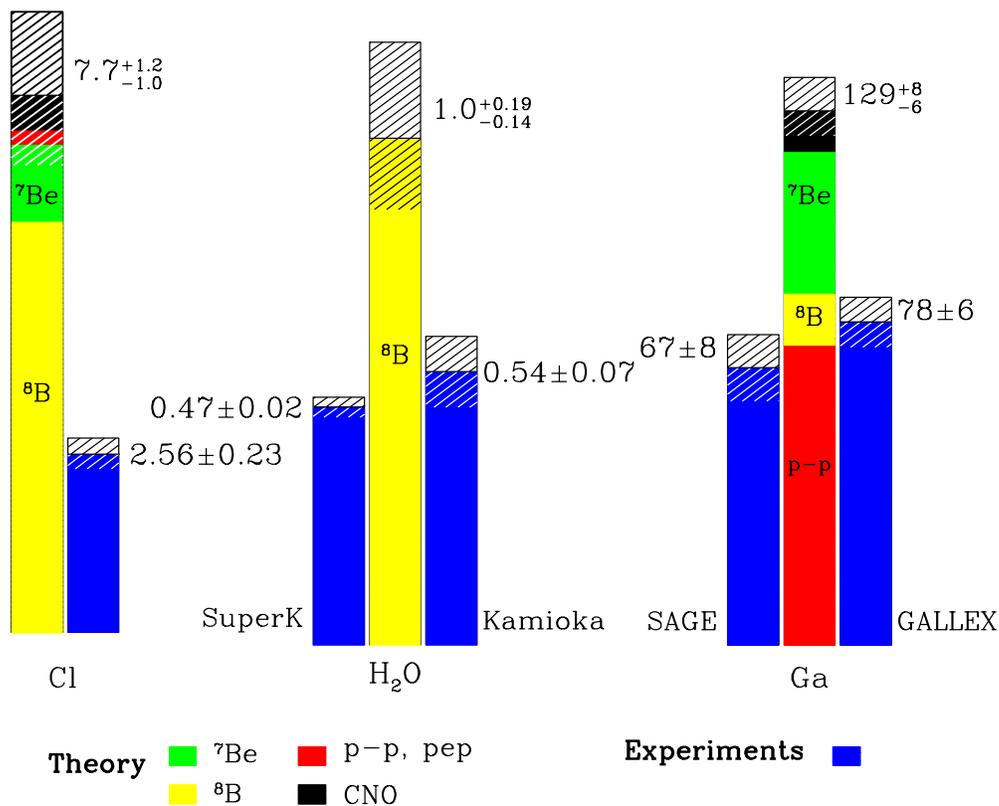,height=18.5cm,angle=270}}
\hss}
\caption{\label{solexpfig}Solar neutrino measurements vs.\ theoretical
flux predictions. (Figure taken from http://www.sns.ias.edu/$\sim$jnb/.)}
\end{figure}

As discussed in Chapter~$1.1$, there are six different solar neutrino
reactions in the $pp$ chain with six different energy spectra.  Different
experiments measure neutrinos from different reactions because they have
different energy thresholds and different spectral response characteristics.
Some of the experiments are Homestake \cite{homestake} which uses the
detection reaction $\nu_e + {\rm ^{37}Cl} \ra {\rm ^{37}Ar} + e^-$ (threshold
814~keV), the gallium experiments GALLEX \cite{gallex}, and SAGE \cite{sage}
which use $\nu_e + {\rm ^{71}Ga} \ra {\rm ^{71}Ge} + e^-$ (threshold 232~keV)
and (Super)Kamiokande \cite{SKsol,kamiokande} where the elastic scattering on
electrons $\en + e^- \ra \en + e^-$ is used (threshold 5~MeV).  These
experiments have in common that they are deep under the Earth (typically some
1000~m water equivalent) to eliminate backgrounds from cosmic radiation.

Soon after the first experiments were started it was found that one half to
two thirds of the neutrino flux was missing.  This ``solar neutrino problem''
is illustrated by Fig.~\ref{solexpfig} where the measured rates of the
chlorine, gallium, and water Cherenkov experiments are juxtaposed with the
predictions in the absence of oscillations.  It is important to note that the
flux suppression is not the same factor in all experiments. Rather, it looks
as if there was a distinct spectral dependence of the neutrino deficit.

In particular, it seems as if the $^7$Be neutrinos do not reach Earth.
Best--fit solutions for SSM flux variations even yield negative $^7$Be fluxes
unless neutrino oscillations are taken into account.  Possible
non--oscillation explanations seem to be unable to explain this scenario in
the light of helioseismology which shows excellent agreement with the SSM.  The
fact that the nuclear cross sections necessary for the SSM are not known for
the relevant (low) energies but have to be extrapolated from higher energy
experiments also can not give the measured rates.  Temperatures different from
the SSM assumption are constrained by the crucial $T$ dependence of the
fluxes, namely $\phi_{\nu} ({\rm B^8}) \propto T^{18}$ and $\phi_{\nu} ({\rm
Be^7}) \propto T^{8}$ which do not allow to reduce the beryllium flux by a
larger factor than the boron flux.

The solar neutrino measurements can be consistently interpreted in terms of
two--flavor oscillations. In contrast to the atmospheric and reactor results,
several different solutions exist as shown in Fig.~\ref{mixing}.  Typical
best--fit points are \cite{solpar}
\be \label{dataused}
\baz
(\delta m^2 \; {\rm (eV^2)}, \; \sin^2 2 \theta ) = & 
\left\{ \baz ( 5.4 \times 10^{-6},\,
6.0 \times 10^{-3}) & \mbox{ SAMSW}  \\[0.2cm]
            ( 1.8 \times 10^{-5},\,0.76)  & \mbox{ LAMSW}\\[0.2cm]
            ( 8.0 \times 10^{-11},\,0.75)  & \mbox{ VO}\\[0.3cm] 
\ea \right. 
\ea 
\ee
Here, SAMSW and LAMSW denote the small--angle and large--angle MSW solution,
respectively, while VO refers to vacuum oscillation.
 
Strategies to decide between these solutions include more precise
investigation of the electron energy spectrum (SAMSW/LAMSW) or seasonal
variations (VO).  A comparison of NC and CC events in the SNO
experiment~\cite{SNO} will clarify if the $\nu_e$ oscillate into sequential or
sterile neutrinos.

\subsection{Summary of Experimental Results} 

With three different masses there can be only two independent $\delta m^2$
values.  Should all experiments with evidence for oscillations be confirmed
then there must be a fourth neutrino, which has to be sterile because LEP
measured the number of sequential neutrinos with $m_{\nu} < 45$ GeV to be~3.
The current results can be summarized as  
\be \label{results}
\bad
(\delta m^2  \; {\rm (eV^2)},\; \sin^2 2 \theta) \simeq & \left\{ \begin{array}{cc} 
(10^{-3} , \ga 0.7) & \mbox{ Atmospheric} \\[0.3cm]
\left.
\begin{array}{cc}
\hspace{2cm} (10^{-5} , 10^{-3})  & \mbox{ SAMSW} \\[0.2cm]
\hspace{2.1cm} (10^{-5} , \ga 0.7)  & \mbox{ LAMSW} \\[0.2cm]
\hspace{2.2cm} (10^{-10} , \ga 0.7) & \mbox{ VO}   \\[0.2cm]
\end{array}  \right\} & \mbox{ Solar} \\[0.3cm] 
(1 ,  10^{-3}) & \mbox{ LSND} \end{array} \right. 
\ea
\ee
Many authors tend to ignore the LSND results to avoid the seemingly unnatural
possibility of a sterile neutrino. On the other hand, if it were found to
exist, after all, this would be the most important discovery in 
neutrino--oscillation physics. 

How to choose the masses to get the observed differences is a topic of its
own. The most interesting aspect of the mass and mixing scheme is the
influence on neutrinoless double beta decay.  We refer to \cite{bilenky99} for
a review of that interesting issue. 
For later need in the next chapter it suffices to say that 
if the atmospheric mass scale is interpreted in a hierarchical mass 
scenario ($\delta m^2 \simeq m^2$) we have one mass eigenstate of about 
0.03 eV\@. 
  
We close this chapter with a few words on future important experiments.  For
solar physics, besides the SNO experiment, the BOREXINO \cite{borex}
experiment will measure the crucial $^7$Be flux to a higher precision than its
predecessors.  MiniBoone \cite{miniboone} is designed to close the LSND/KARMEN
debate and thus confirm or refute the need for a sterile neutrino.

Also planned are so--called long baseline (LBL) accelerator experiments from
KEK to Kamioka (K2K, \cite{K2K}), from Fermilab to Soudan (MINOS,
\cite{MINOS}) and from CERN to Gran Sasso (ICARUS, \cite{ICARUS}) probing mass
ranges down to $\delta m^2 \simeq 10^{-3}$ eV and testing possible CP violation
in the neutrino sector.  In the years to come, exciting discoveries are to be
expected.


\newpage

\section{Neutrinos in Cosmology}

\subsection{Friedmann Equation and Cosmological Basics}

The neutrino plus anti--neutrino density per family in the universe of
$113~{\rm cm}^{-3}$ is comparable to the photon density of $411~{\rm
cm}^{-3}$. Therefore, it is no surprise that neutrinos, especially if they
have a non--vanishing mass, may be important in cosmology. To appreciate their
cosmological role, we first need to discuss some basic properties of the
universe.
 
On average, the universe is homogeneous and isotropic. Its expansion is
governed by the Friedmann equation, 
\be \label{friedmann}
H^2 = \frac{8 \pi}{3} G_N \rho - \frac{k}{a^2}, 
\ee
where $G_N$ is Newton's constant, $\rho$ the energy density of the universe,
and $H = \dot{a}/a$ the expansion parameter with $a(t)$ the cosmic scale
factor with the dimension of a length. The present--day value of the
expansion parameter is usually called the Hubble constant. It has the
value
\be
H_0=h\,100~\rm km~s^{-1}~Mpc^{-1},
\ee
where $h=0.5$--0.8 is a dimensionless ``fudge factor.''    

The constant $k$ determines the spatial geometry of the universe which is
Euclidean (flat) for $k=0$, and positively or negatively curved with radius
$a$ for $k=\pm1$, respectively. The universe is spatially closed and will
recollapse in the future for $k=+1$. For $k=0$ or $-1$ it expands forever and
is spatially infinite (open), assuming the simplest topological structure.
However, even a flat or negatively curved universe can be closed. An example
for a flat closed geometry is a periodic space, i.e.\ one with the topology of
a torus.

Equation~(\ref{friedmann}) must be derived from Einstein's field equations.
However, it can also be heuristically derived by a simple Newtonian argument.
Since the universe is assumed to be isotropic about every point, we may pick
one arbitrary center as the origin of a coordinate system.  Next, we consider
a test mass $m$ at a distance $R(t) = a(t)\,r$ from the center, and assume that
the homogeneous gravitating mass density is $\rho$.  The total energy of the
test mass is conserved and may be written as $E = -\frac{1}{2} K m$.
On the other hand,   
\be
E = T + V = \frac{1}{2} m \dot{R}^2 - \frac{G_N M m}{R} ,  
\ee
where $M = R^3 \rho\, 4 \pi/3$ is the total mass enclosed by a sphere of radius
$R$. With $K = k r^2$ the Friedmann equation follows. 
On the basis of a Galileo transformation one can easily show that the
Friedmann equation stays the same when transformed to another point, i.e.\
the expansion looks isotropic from every chosen center. 

Some basic characteristics of the expanding universe are easily understood if
we study the scaling behavior of the energy density $\rho$. Nonrelativistic
matter (``dust'') is simply diluted by the expansion, while the total number
of ``particles'' in a co--moving volume, of course, is conserved.  Therefore,
when matter dominates, we find $\rho\propto a^{-3}$.

For radiation (massless particles), the total number in a comoving volume is
also conserved. In addition, we must take the redshift by the cosmic expansion
into account. The simplest heuristic derivation is to observe that the cosmic
expansion stretches space like a rubber--sheet and thus stretches the periodic
pattern defined by a wave phenomenon. Thus the wavelength $\lambda$ of a
particle grows with the cosmic scale factor $a$, implying that its wave number
$k$ and thus its momentum scale as $a^{-1}$. For radiation we have $\omega=k$
so that the energy of every quantum of radiation decreases inversely with the
cosmic scale factor. In summary, the energy density of radiation scales as
$a^{-4}$.

For thermal radiation (blackbody radiation) we may employ the Stefan--Boltzmann
law $\rho \propto T^4$ to see that $T\propto a^{-1}$. The temperature of the
cosmic microwave background radiation is a direct proxy for the inverse
of the cosmic scale factor. 

The density in Friedmann's equation, therefore, decreases at least with
$a^{-3}$ so that at late times the curvature term takes over if $k=\pm1$.  Of
course, it is frequently assumed that the universe is flat, and certainly the
curvature term, if present at all, may only begin to be important today. For
$k=-1$, $H^2$ will never change sign, so the universe expands forever.  For
$k=+1$, the expansion turns around when $H^2=0$, i.e.\ when $a^{-2}=(8\pi/3)
G_N\rho$.  Either way, at very early times (large temperature, small scale
factor), radiation dominates. Therefore, in the early universe we may neglect
the curvature term.

A ``critical'' or Euclidean universe is characterized by $k = 0$. In this case
we have a unique relationship between $\rho$ and $H$. The density corresponding
to the present--day expansion parameter
$H_0$ is called the critical density   
\be
\rho_c = \frac{3 H_0^2}{8 \pi G_N} =h^2\,
1.88 \times 10^{-29}\rm~g~cm^{-3} .
\ee
The last number is the current value of the critical density, denoted 
$\rho_0$. It translates to about $10^{-5}$ protons per cm$^{3}$. 
The density is
often expressed as a fraction of the critical density by virtue of
$\Omega\equiv\rho/\rho_c$.  A value of $\Omega=1$ corresponds to a flat
universe.

The cosmic background radiation, which was first observed in the 1960s, is
equivalent to the radiation of a black body with temperature $T = 2.726$~K.
Therefore, today's number density of photons is
\be
n_{\gamma} = 2 \int \frac{d^3{\bf p}}{(2 \pi)^3}\, f (\omega)
= \frac{2 \zeta_3}{\pi^2}\, T^3  \simeq 411.5~{\rm cm^{-3}} ,  
\ee
where $f(\omega)=(e^{\omega/T} - 1)^{-1}$ is the Bose--Einstein distribution
function.  The energy density of photons (or in general bosons) in the
universe is calculated by
\be \label{rhogamma}
\rho_{\gamma} = 2 \int \frac{d^3 {\bf p}}{(2 \pi)^3} \omega f (\omega)
= \frac{\pi^2}{15}\, T^4 \equiv g_{\rm B}\,  \frac{\pi^2}{30}\, T^4 ,  
\ee 
where we have introduced the effective number of boson degrees of freedom
$g_{\rm B}$ for later use. Therefore,
\be
\Omega_{\gamma} \equiv 
\frac{\rho_{\gamma 0}}{\rho_0} \simeq 
4.658 \times 10^{-34}~{\rm g~cm^{-3}}/\rho_0  \simeq 
2.480 \times 10^{-5}  \, h^{-2} 
\ee
is the contribution of microwave photons to the critical density.

The fraction of baryons relative to the number of photons in the 
present--day universe is   
\be
\eta = \frac{n_B}{n_{\gamma}} = \eta_{10} \times 10^{-10} 
\simeq  3 \times 10^{-10} , 
\ee 
so that the contribution of baryons to the critical density is 
merely  
\be \label{omegab} 
\Omega_{\rm B} \equiv  
\frac{\rho_{\rm B 0} }{\rho_0} = 
\frac{\eta \langle E_B \rangle n_{\gamma}}{\rho_0} 
= \frac{\eta m_p n_{\gamma}}{\rho_0} 
\simeq 3.6 \times 10^{-3} \, h^{-2} \, \eta_{10}
\simeq 0.01 \, h^{-2} , 
\ee 
where we took the proton mass as an average baryonic energy.

\subsection{Radiation Epoch}

In the hot early universe, neutrinos should have been in thermal equilibrium
so that one expects a cosmological neutrino sea in analogy to the cosmic
microwave background. The number density of one massless 
thermal neutrino generation is 
\be \label{ferminb}
n_{\nu\bar\nu} = 2 \int \frac{d^3{\bf p}}{(2 \pi)^3} f(\omega) 
=  \frac{3 \zeta_3}{2 \pi^2} T^3 , 
\ee
where we have to use the Fermi--Dirac distribution
$f(\omega)=(e^{\omega/T}+1)^{-1}$ for fermions with a vanishing chemical
potential. Equation~(\ref{ferminb}) counts two internal degrees of freedom,
i.e.\ it counts left--handed (interacting) neutrinos and 
(right--handed) anti--neutrinos for
a given family. The energy density is
\be \label{fermirho}
\rho_{\nu\bar\nu} = 
g_{\rm F} \int \frac{d^3 {\bf p}}{(2 \pi)^3}\,\omega f(\omega) 
= g_{\rm F}\,\frac{7}{8}\, \frac{\pi^2}{30}\, T^4 
\ee
with the effective number of fermion degrees of freedom $g_{\rm F}$. The
effective number of thermal degrees of freedom is defined by
\be
g_{\ast} = \sum_{j=\rm bosons} g_j + 
\frac{7}{8} \sum_{j=\rm fermions} g_j, 
\ee
so that the total energy density in radiation is
\be \label{eq:rhorad}
\rho_{\rm rad}=g_* \frac{\pi^2}{30}\,T^4.
\ee
Another important quantity is the entropy density in radiation
$S = (p + \rho)/T$. Since for a relativistic gas $p = \rho/3$ 
we have
\be\label{eq:entropy}
S = g_*\,\frac{2\pi^2}{45}\,T^3 . 
\ee
The energy and entropy densities are simply functions of the temperature and
of the particle degrees of freedom that are thermally excited at that
temperature.

To get a feeling for $g_{\ast}$, we consider some characteristic temperatures.
At $T<100$~MeV, below the QCD phase transition, there are photons, $e^{\pm}$,
and three flavors of (anti)neutrinos, giving $g_{\ast} = 2 + 7/8 \times 10 =
43/4=10.75$.  At $T=300$~MeV, significantly above the QCD phase transition, in
addition muons (4 fermionic degrees of freedom), eight gluons ($2\times8$
bosonic degrees of freedom) and up, down, and strange quarks
($3\times4\times3$ fermionic degrees of freedom, counting 3 color degrees of
freedom each) were present, resulting in $g_*= 18 + 7/8\times 44 = 113/2=56.5$.
The number of degrees of freedom drops dramatically around the QCD phase
transition when quarks and gluons condense to confined states of mesons and
baryons.

The Friedmann equation for the radiation epoch in the early
universe can be written in the form
\be \label{friedfundb}
H^2=\frac{4 \pi^3}{45} G_N g_{\ast}T^4. 
\ee
We can now calculate a relationship between cosmic age and temperature in
the radiation epoch. We use $\rho = \rho_0 (R_0/R)^4 \equiv b R^{-4}$ and 
$H^2 = 8 \pi/3 G_N b R^{-4}$ which leads to 
\be 
\dot{R}(t) = \left(\frac{8 \pi}{3} G_N b\right)^{1/2}\, \frac{1}{R}
\ee
and thus to
\be
R(t) = \left( \frac{32 \pi}{3} G_N b \right)^{1/4} t^{1/2} . 
\ee
With Eq.\ (\ref{eq:rhorad}) for $\rho(T)$ we find 
\be
T = \left( \frac{45}{16 \pi^3 G_N g_{\ast}} \right)^{1/4} t^{-1/2}
= g_{\ast}^{-1/4} \, 1.56~{\rm MeV}\left(\frac{1~\rm s}{t}\right)^{1/2}. 
\ee
Therefore, the universe had a temperature of about 1~MeV when it was about
1~s old. 

\subsection{Present--Day Neutrino Density}
 
When neutrinos and photons are in thermal equilibrium in the early universe,
they are both characterized by the same temperature. However, the neutrino
temperature today is thought to be lower than that of the cosmic microwave
background. The reason for the difference is that 
when the temperature has fallen below the electron
mass $m_e$ ($t \simeq  10$ s) the annihilation of
electrons and positrons heats the photon gas by virtue of
$e^+e^-\ra 2\gamma$. 
As we will see later, neutrinos have already decoupled at
that time  so that the rate for $e^+e^-\ra \nu\bar\nu$ is
too slow to be of importance.

The electron--positron--photon plasma is so tightly coupled that one may assume
that it is always close to thermal equilibrium throughout the annihilation
process, i.e.\ the disappearance of the $e^\pm$ pairs is an adiabatic process.
Therefore, the entropy of the $e^{\pm}$ gas will go over to the photons. If we
denote the photon temperature before this process as $T_1$ and afterward as
$T_2$, and if the value of $g_*$ for the coupled species before and after
annihilation is $g_{1,2}$, respectively, entropy conservation implies
\be
g_1T_1^3=g_2T_2^3.
\ee
Before annihilation, the participating species are $e^\pm$ and photons so that
$g_*=2+4\times 7/8=11/2$, while afterward we have only photons with
$g_*=2$. Before annihilation, we have $T_\nu=T_\gamma$ so that the
ratio $T_2/T_1$ can be interpreted as $T_\gamma/T_\nu$ after annihilation. 
Therefore, we find the famous result
\be
T_{\nu} = \left(\frac{4}{11}\right)^{1/3} T_{\gamma} . 
\ee
Thus, the present--day temperature of the neutrino background is
predicted to be $T_{\nu} \simeq 1.946~\rm K \simeq 1.678 \times 10^{-4}$~eV\@.
Naturally, its direct detection is an impossible task.
 
The number density of one neutrino familiy is 
given by Eq.~(\ref{ferminb}). Inserting the above numbers yields 
\be \label{numu} 
n_{\nu\bar\nu} \simeq 337.5 \, \rm cm^{-3} 
\ee 
for all three generations. 
If all neutrinos are relativistic today, all three families 
contribute 
\be
\rho_{\nu\bar\nu} = \frac{7}{8} \left( \frac{4}{11} \right)^{4/3} 
\frac{3 g_{\nu}}{g_{\gamma}} \rho_{\gamma} 
\simeq 3.174 \times 10^{-34} \rm~g~cm^{-3}  
\ee
to the cosmic energy density.

We have assumed that the neutrinos are relativistic, which is the case when 
their mass is smaller than their average energy 
\be
\langle E_{\nu} \rangle  = \frac{\rho_{\nu\bar\nu}}{n_{\nu\bar\nu}}
\simeq 3 \, T_\nu
\simeq 
5 \times 10^{-4}~\rm eV. 
\ee
If the indications for neutrino oscillations discussed in the previous
section are correct, at least one of the mass eigenstates exceeds around
0.03~eV and thus is not relativistic today.

But even if $m_{\nu}$ is bigger than $ 5 \times 10^{-4}$~eV, the number
density is still given by Eq.~(\ref{numu}) since at the point of $e^\pm$
annihilation the neutrinos were relativistic. 
The energy density today, though,
changes for one nonrelativistic flavor to $\rho_{\nu\bar\nu} = n_{\nu\bar\nu}
m_{\nu}$.  The contribution to the critical density today is then
$\Omega_{\nu\bar\nu} \equiv \rho_{\nu\bar\nu}/\rho_c$ or
\be
\Omega_{\nu\bar\nu}h^2 \simeq 
\sum_{\rm flavors} \frac{m_{\nu}}{94~\rm eV}. 
\ee
We may use this result to derive the famous cosmological limit on the
neutrino masses. It turns out that the lower limit on the age of the universe
indicated by the age of globular clusters implies something like 
$\Omega h^2 \stackrel{<}{\sim} 0.4$. Since $\Omega_{\nu\bar\nu} < \Omega$ 
we have 
\be
\sum_{\rm flavors} m_\nu \stackrel{<}{\sim} 37~\rm eV.    
\ee
If the experimental indications for neutrino oscillations are correct, the
mass differences are very small. Therefore, 
a neutrino mass in the neighborhood of this limit would require that all
flavors have nearly degenerate masses, implying that we may divide this
limit by the number of flavors to obtain a limit on the individual masses.
Therefore,
\be\label{eq:cosmicmasslimit}
m_\nu\stackrel{<}{\sim} 12~\rm eV
\ee
applies to all sequential neutrinos. 

If the current indications for oscillations from the atmospheric neutrino
anomaly are correct so that one mass eigenstate exceeds about 0.03~eV we find
that $\Omega_{\nu\bar\nu}h^2\stackrel{>}{\sim}3\times10^{-4}$ or, with
$h\stackrel{<}{\sim}0.8$, we have
$\Omega_{\nu\bar\nu}\stackrel{>}{\sim}5\times10^{-4}$, not much less than the
luminous matter of the universe.

One may wonder if any of these results change if neutrinos are Dirac particles
so that there are actually 4 degrees of freedom per flavor. However, the
sterile (right--handed) components will not be thermally excited in the
relevant epochs of the early universe because they are too weakly coupled if
the masses are as small as indicated by the cosmological limit. It is
important to realize that $g_*$ is the effective number of {\it thermally
  excited\/} degrees of freedom, which does not necessarily include all
existing particles. For example, the massless gravitons are another species
which are not thermally excited anywhere near the QCD or $e^\pm$ annihilation
epochs so that they, too, have not appeared in our particle counting for
$g_*$.

\subsection{Big Bang Nucleosynthesis}

How do we know that the cosmological neutrino background actually exists?
After all, its direct detection is an impossible task with present--day
experimental means. However, the cosmic neutrinos manifest themselves directly
in the process of forming the lightest nuclei in the early universe so that
Big Bang Nucleosynthesis (BBN) yields compelling evidence for the reality of
the ``cosmic neutrino sea.''

All of the deuterium and most of the helium in the present--day universe can
not have been produced in stars, so that it must have been produced shortly
after the Big Bang.  The process of primordial nucleosynthesis explains
today's abundance of several light elements, notably $^4$He, D, $^3$He and
$^7$Li.  Perhaps the most important quantity is the mass fraction $Y_p$ of
primordial $^4$He, which is observationally inferred to be 
\be 
Y_p \equiv
\frac{M_{\rm He}}{ M_{\rm H} + M_{\rm He}} \simeq 0.24 .  
\ee 
The abundances of the other elements are very much smaller; for example
$n_{\rm D + ^3He}/n_{\rm H} \simeq 10^{-5}$, $n_{\rm ^7Li}/n_{\rm H} \simeq
10^{-10}$.

In a thermal plasma, all nuclei should be present with their thermal
equilibrium abundance given by nuclear statistical equilibrium (NSE). 
The abundance of nuclei with atomic number $A$, nuclear charge $Z$, 
binding energy $B_A$, and a statistical factor $g_A$ is~\cite{kolbturner} 
\be
n_A = g_A A^{3/2} 2^{-A} \left( \frac{2 \pi}{m_N T} 
\right)^{\frac{3 (A - 1)}{2}} n_p^Z n_N^{A-Z} \exp{B_A/T} . 
\ee
For example, $g_A =2$ for $^3$He which has total spin 1/2. 

At high temperatures, there are very few nuclei because the dissociated state
is preferred. The early universe is a ``high--entropy environment'' because
there are about $10^{10}$ thermal photons per baryon. Therefore, the
dissociated state is favored until very late when the temperature is
significantly below the typical nuclear binding energies. The main idea behind
BBN is that nuclei stayed dissociated until very late so that they appear in
appreciable numbers only at a time when the reactions forming and dissociating
them begin to freeze out of thermal equilibrium. Therefore, hardly any nuclei
heavier than helium form at all ---otherwise one might 
have expected that most
of the matter appears in the form of iron, the most tightly bound nucleus.

Nuclei are produced from protons and neutrons which, at high
temperatures, are in thermal equilibrium by $\beta$--processes of the form
\be \label{reactions}
\bar\nu_e + p \lra e^+ + n, \qquad \en + n \lra e^- + p, \qquad 
n \lra p + e^- + \bar\nu_e. 
\ee
These processes ``freeze out'' when they become slow relative to the expansion
rate of the universe. By dimensional analysis the reaction rates are of the form
\be
\Gamma \propto G_F^2 T^5 . 
\ee
The other characteristic time scale is defined by the Hubble parameter $H$
which has the form
\be
H \propto \frac{1}{m_{\rm Pl}} T^2 ,  
\ee 
where we write Newton's constant as the inverse of the Planck mass 
squared. Therefore the expansion of the universe was too fast for the 
rates when $H > \Gamma$, which leads to a freeze--out temperature of 
\be 
T_F \simeq 0.8~\rm MeV. 
\ee 
At lower temperatures (later times) weak processes and thus neutrinos are
no longer in equilibrium. 

Before the weak reactions became ineffective, the ratio of protons and
neutrons was given thermal equilibrium as 
\be 
\frac{n_n}{n_p} = \exp \left(\frac{-Q}{T}\right), 
\ee 
where $Q=m_n-m_p=1.293$~MeV.  When $T < T_F$ (after about $1$~s), the ratio
changes only by neutron free decays with a neutron lifetime of $\tau_n \simeq
887$~s, i.e.\ it decreases as $\exp \left(-t/\tau_n \right)$.

At $T_F$ when the weak interactions froze out, the neutron/proton ratio was
about $1/6$.  After the temperature fell below about $0.1$~MeV at a cosmic age
of about 3~min, BBN began in earnest. Some important reactions involving the
first few steps are  
\be 
\bad 
& n + p \lra \Deu + \gamma  \; \, ({\rm E_B = 2.2 \, MeV}),  & \\
\Deu + p \lra { \rm^3He} + n , & \Deu + n \lra { \rm^3H} + \gamma , & 
\Deu + \Deu \lra { \rm^3He} + n , \\
{ \rm^3He} + n \lra { \rm^4He} + \gamma , 
& { \rm^3H} + p \lra { \rm^4He} + \gamma  , 
& \Deu + \Deu \lra  { \rm^4He} + \gamma. 
\ea
\ee 
Most of the reactions end in $\rm ^4He$. Heavier elements do not efficiently
form because they have too large Coloumb barriers. Moreover, there are
bottlenecks in the reaction network because there are no stable nuclei with $A
= 5$ or $A = 8$.

\begin{figure}[hp] 
\setlength{\unitlength}{1cm}
\begin{center}
\epsfig{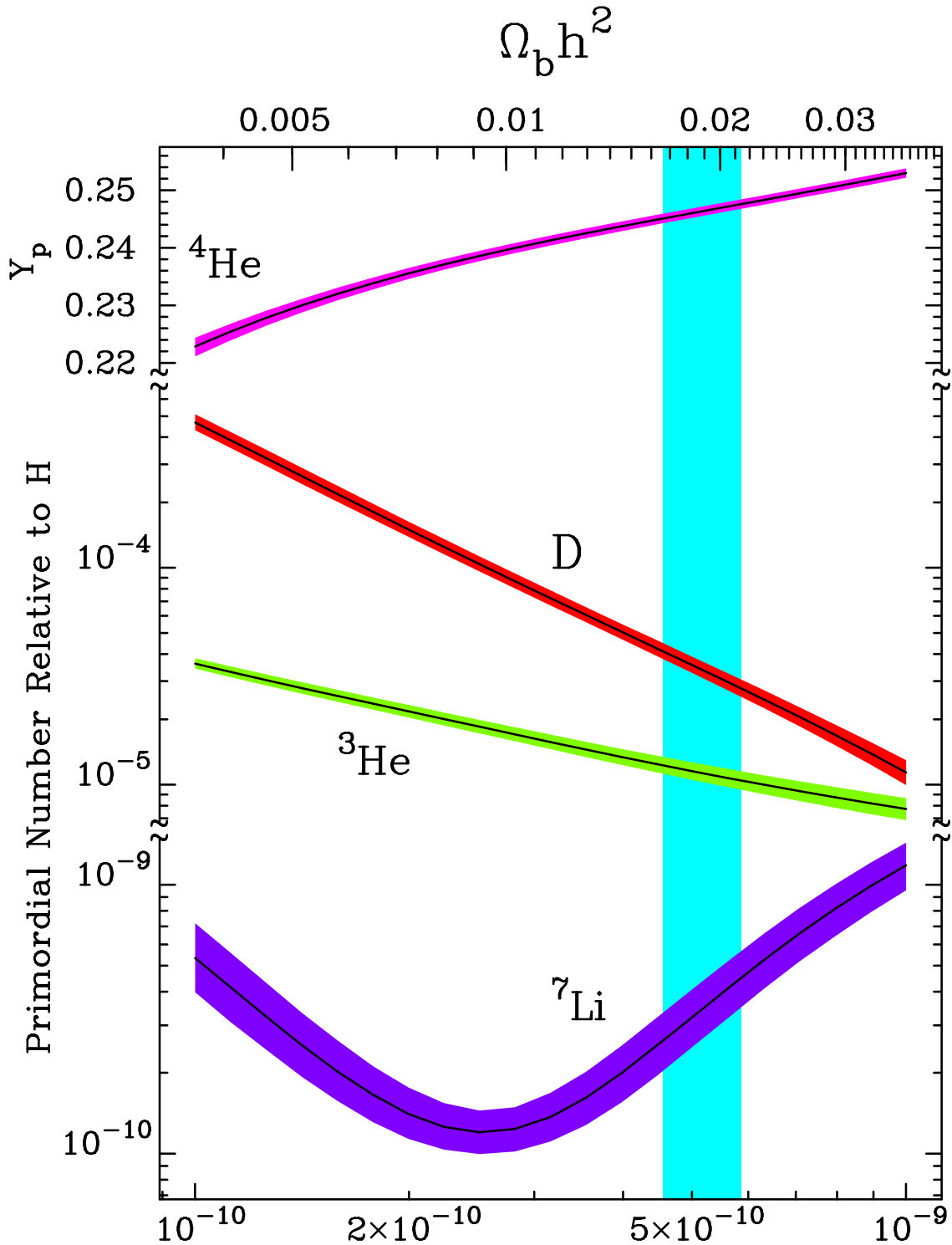}
\vspace{-1.0cm}
\caption{\label{etafig}Typical BBN predictions (solid line) 
for $Y_p$, D,  $\rm ^3He$ and $\rm ^7Li$ relative to $H$ with 
$2\sigma$ theoretical uncertainty. The vertical band 
represents a recent 
measurement of the deuterium abundance \cite{BurTyt}. 
(Figure from \cite{Bur}.)}
\end{center}
\end{figure}

At higher temperatures the deuteron produced in the first reaction was
immediately dissociated by photons, which have an average energy of $\langle
E_{\gamma} \rangle = \rho_\gamma / n_\gamma \simeq 2.70 \, T$.  
These dissociating photons were available in
great number since the baryon/photon ratio $\eta = \eta_{10} 10^{-10}$ is very
small. When $\eta \exp \left(E_B/T\right) \simeq 1$, deuterons were produced
in sufficient numbers. This requirement gives a temperature of about 0.1~MeV
and thus defines the beginning of~BBN.

When all remaining neutrons had disappeared in $\rm ^4He$, we have $M_{\rm He}
= n_n 2 m_N$ and $M_{\rm H} = (n_p - n_n) m_N$, which leads to a BBN
prediction for the helium mass fraction of
\be
Y_p = \frac{2 n_n}{n_p + n_n} . 
\ee
Since BBN stopped about three minutes after the weak reactions
Eq.~(\ref{reactions}) froze out, we get $n_n/n_p \simeq 1/7$ and thus $Y_p
\simeq 0.24$ in agreement with the observations.

All of our estimates depend on the freeze--out temperature $T_F$. It would be
higher if $H$ were larger than implied by the standard value for $g_*$.  A
larger $T_F$ implies an increased $n_n/n_p$ and thus an increase of $Y_p$.
Since $H$ is proportional to $g_{\ast}^{1/2}$ we find that additional neutrino
families would lead to a higher $Y_p$. This argument was used to constrain the
number of neutrino flavors before the famous LEP experiment at
CERN. In \cite{Bur} one finds that for $Y_p = 0.246 \pm 0.0014$ the number 
of neutrino species is limited by $N_{\nu} < 3.20 $ at 95\%~C.L. 

The theoretical dependence of the light element abundances on $\eta$ is shown
in Fig.~\ref{etafig}, assuming the standard number of neutrino families
$N_\nu=3$. The BBN predictions explain abundances consistently over a range of
10 orders of magnitude. Therefore, at the present time it appears that BBN is
a consistent theory. It certainly requires the presence of thermal neutrinos
and thus can be taken as compelling evidence for the reality of the cosmic
neutrino sea.

\subsection{Neutrinos as Dark Matter}

The experimental evidence for neutrino oscillations is now so compelling that
the reality of non--vanishing neutrino masses must be taken as a serious
working hypothesis. However, neutrinos are not a good candidate for the
ubiquitous dark matter (DM) which dominates the dynamics of the universe.

One of the most conspicuous dark matter problems is that of the flat 
galactic rotation curves, implying that the dynamical
mass of the galaxy is dominated by some non--luminous component. Assuming that
this mass density consist of neutrinos, their maximum phase--space density 
implied by the Pauli exclusion principle is 
\be
n_{\rm max} \equiv \frac{p_{\rm max}^3}{3 \pi^2} . 
\ee 
Here, $p_{\rm max} = \mn v_{\rm max}$ is the maximum momentum of a neutrino
bound to the galaxy with $v_{\rm max}$ the escape velocity of order
$500\rm~km~s^{-1}$. Since $\rho_{\nu \rm DM} = n_{\nu} \mn$ we obtain a {\it
  lower\/} limit for $\mn$. For typical spiral galaxies one finds
$m_\nu>20$--30~eV, and much larger values for dwarf galaxies which are
dominated by dark matter.  This limit, known as the Tremaine--Gunn bound
\cite{TreGun}, at least nominally excludes neutrinos as the main component of
galactic dark matter. 

Even more severe problems arise from arguments about cosmic structure
formation. The universe is thought to have started from an almost homogeneous
early phase with low--amplitude primordial density fluctuations which must have
been produced by some physical process; one favored scenario involves the
generation of the primordial fluctuation spectrum during an early phase of
exponential expansion (inflation). The density contrast of these primordial
fluctuations increases by the action of gravity since any region with more
mass than its surroundings will attract more mass, at the expense of
lower--density regions. This gravitational instability mechanism can
beautifully account for the observed structure in the matter distribution.

However, weakly interacting particles become collisionless early and
thus can travel large distances undisturbed. In this way they will
wash out the primordial density fields by their ``free streaming'' or
``collisionless phase mixing'' and thus remove the seeds for structure
formation up to a certain scale. The smaller the mass, the later these
particles will become nonrelativistic, the further they travel, and
thus the larger the scales below which the primordial field of density
fluctuations has been erased. If all scales up to those which later
correspond to galaxies have been erased, the particles are called
``hot dark matter'' while otherwise they are referred to as ``cold
dark matter.'' The division line between the two notions corresponds
to a particle mass in the few keV range.

Since hot dark matter can not account for the observed structures
within the standard model of structure formation, they are strongly
disfavored, except perhaps as a subdominant component in a hot plus
cold dark matter scenario. It is quite surprising that oscillation
experiments have established neutrino masses, but that neutrinos
nevertheless are no good dark matter candidates.

\begin{figure}[ht] 
\setlength{\unitlength}{1cm}
\begin{center}
\epsfig{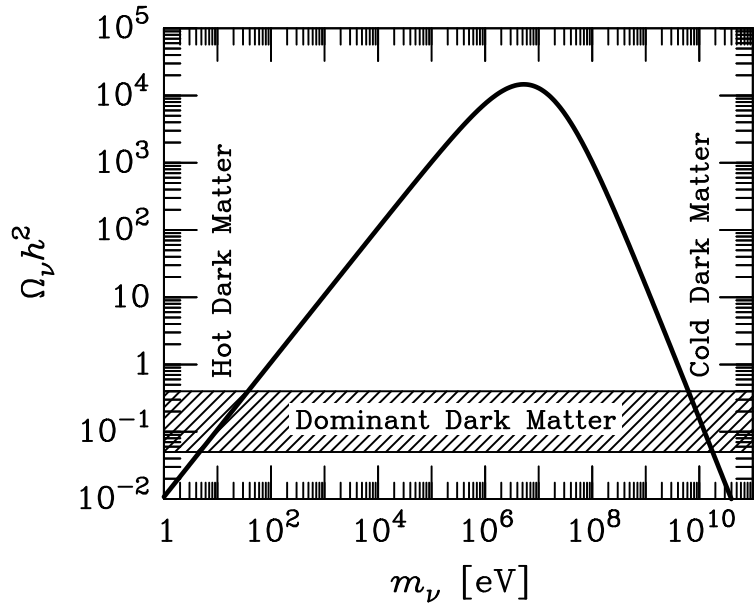}
\caption{\label{neucon}Contribution of massive neutrinos to $\Omega$.}
\end{center}
\end{figure}

It is widely believed that the dark matter consists of some new,
heavy, neutrino--like particles generically referred to as WIMPs (weakly
interacting massive particles). One may think that such a cold dark matter
particle would vastly overclose the universe since the cosmological mass limit
on neutrinos requires a mass less than a few ten eV. However, if the
weakly interacting particles are heavier than a few MeV, the weak--interaction
freeze--out happens when the temperature has fallen below their mass. 
Therefore, their number density is decimated by annihilations, i.e.\ their
thermal equilibrium density is suppressed by a Boltzmann factor
$e^{-m/T}$ until the annihilation rate becomes too slow to keep up with the
cosmic expansion. Therefore, as a function of the assumed mass of a 
weakly interacting particle one obtains a cosmic density contribution
schematically shown in Fig.~\ref{neucon}. This ``Lee--Weinberg curve''
turns over for large neutrino masses so that there is a second solution
at masses of a few or a few ten GeV where neutrinos could be the dark matter,
and then of the cold variety.

Of course, none of the sequential neutrinos can play this role because their
masses are too small. Even if we ignore the oscillation data, the direct
kinematical mass limits are so restrictive (about 18~MeV for $\nu_\tau$, the
worst case) that standard neutrinos can not play the role of cold dark matter.
However, in popular extensions of the standard model the existence of the
requisite particles is quite plausible.  Notably the theory of supersymmetry
provides an ideal candidate in the guise of the ``neutralino,'' a neutral
spin--1/2 Majorana fermion. From a cosmological perspective, these neutralinos
are virtually identical with a heavy Majorana neutrino. Even if the standard
neutrinos are not the dark matter of the universe, it still looks like weakly
interacting particles make up the bulk of the mass in our universe.


\newpage

\section{Conclusions}

The question if neutrinos, the most elusive of all elementary particles, have
nonvanishing masses has come very close to a conclusion over the past few
years. For a long time it had been suspected that the solar neutrino problem
is solved by neutrino oscillations and thus by neutrino masses. However, it
took the spectacular up--down asymmetry of the atmospheric neutrino flux
observed by the SuperKamiokande detector to convince most of the sceptics that
neutrino oscillations are real. It is remarkable that natural neutrino
sources, the Sun and the upper atmosphere, play a central role at putting
together the pieces of the jigsaw puzzle of the leptonic CKM matrix.

As we learn more about the neutrino masses and mixing parameters, there is a
certain sense that neutrino masses are too small to be of great cosmological
importance. To be sure, it is still possible that the global neutrino mass
scale is much larger than their mass differences. In such a degenerate scheme,
neutrinos could still play a role as a subdominant dark matter component in
hot plus cold dark matter cosmologies. It is also still possible that sterile
neutrinos exist. If this hypothesis were verified, neutrinos would be assured
of an important role for cosmic structure formation and might modify the
standard theory of Big Bang nucleosynthesis.

One should take note that neutrino astrophysics is a field much broader than
the question of the neutrino mixing matrix. Once the mixing parameters and
masses have been settled, one has greater confidence in one's treatment of
astrophysical phenomena were neutrino masses and oscillations are potentially
important such as supernovae or gamma--ray bursts, or in the use of neutrinos
as a new form of radiation to do astronomy with. Neutrino physics and
astrophysics will keep us busy for some time to come!


\newpage

\setcounter{section}{0}
\renewcommand{\thesection}{\Alph{section}}
\def\chaptername{Appendix}
\setcounter{table}{0}

\section{\label{tables}Useful Tables}
\subsection{\label{integrale}Integrals}
The following integrals frequently appear in the context of calculations
involving particle reactions in thermal media, where $\zeta$ refers to the
Riemann zeta function.

\begin{table}[bh]
\begin{center}
\caption{Thermal integrals.}
\medskip
\begin{tabular}{cccc} \hline \hline
 & Maxwell--Boltzmann& Fermi--Dirac& Bose--Einstein \\[4pt]
 & $ \D\int \limits_0^{\infty} \frac{x^n}{e^x} dx$ & 
   $\D\int\limits_0^{\infty} \frac{x^n}{e^x + 1} dx$ &
   $\D\int\limits_0^{\infty} \frac{x^n}{e^x - 1} dx$ \\
\noalign{\vskip4pt}
\hline
\noalign{\vskip4pt}
$n = 2$ & 2 & $\frac{3}{2} \zeta_3 \simeq 1.8031 $ & 
$ 2 \zeta_3 \simeq 2.40411 $ \\[6pt]
$ n = 3$ & 6 & $\frac{7 \pi^4}{120} \simeq 5.6822 $ & 
$ \frac{\pi^4}{15} \simeq 6.4939 $ \\[4pt] \hline
\end{tabular}
\end{center}
\end{table}

\subsection{Conversion of Units}

We always use natural units where $\hbar=c=k_{\rm B}=1$. In order to convert
between different measures of length, time, mass, or energy one may use the
following table. For example, 1~cm$^{-1}= 2.998 \times 10^{10}~\rm s^{-1}$.
The atomic mass unit is denoted by amu.

\begin{table}[hbt]
\caption{\label{conversion}Conversion factors for natural units.}
\bigskip
\hbox to\hsize{\hss
\small
\begin{tabular}{|c||c|c|c|c|c|c|c|} \hline 
 & $\rm s^{-1}$ & $\rm cm^{-1}$ & K & eV & amu & erg & g \\ \hline \hline 
$\rm s^{-1}$ & 1 & $0.334{\times}10^{-10} $ & $0.764{\times}10^{-11} $ 
& $0.658{\times}10^{-15} $ &  $0.707{\times}10^{-24} $ & 
$1.055{\times}10^{-27} $ & $1.173{\times}10^{-48}$ \\ \hline
$\rm cm^{-1}$ &  $2.998 {\times} 10^{10}$ & 1 & $0.229                $ 
& $1.973 {\times} 10^{- 5} $ &  $2.118 {\times} 10^{-14} $ 
& $3.161 {\times} 10^{-17} $ 
& $0.352 {\times} 10^{-37}$ \\ \hline
K             &  $1.310 {\times} 10^{11}$ & 4.369 & $1                    $ 
& $0.862 {\times} 10^{- 4} $ &  $0.962 {\times} 10^{-13} $ & 
$1.381 {\times} 10^{-16} $ 
& $1.537 {\times} 10^{-37}$ \\ \hline
eV           &  $1.519 {\times} 10^{15}$ & $0.507 {\times} 10^{5}$ & 
$1.160 {\times} 10^4$ & $1                    $ &  $1.074 {\times} 10^{-9 } $ 
& $1.602 {\times} 10^{-12} $ & $1.783 {\times} 10^{-33}$ \\ \hline
amu        &  $1.415 {\times} 10^{24}$ & $0.472 {\times} 10^{14}$ & 
$1.081 {\times} 10^{13}$ & $0.931 {\times} 10^9     $ &  $1    $ 
& $1.492 {\times} 10^{-3 } $ & $1.661 {\times} 10^{-24}$ \\ \hline
erg     &  $0.948 {\times} 10^{27}$ & $0.316 {\times} 10^{17}$ & 
$0.724 {\times} 10^{16}$ & $0.624 {\times} 10^{12}  $ &  $0.670 {\times} 10^3$ 
& 1                       & $1.113 {\times} 10^{-21}$ \\ \hline
  g     &  $0.852 {\times} 10^{48}$ & $2.843 {\times} 10^{37}$ & 
$0.651 {\times} 10^{37}$ & $0.561 {\times} 10^{33}  $ 
&  $0.602 {\times} 10^{24}  $ 
& $0.899 {\times} 10^{21}$  &  1 \\ \hline
\end{tabular}\hss}
\end{table}


\newpage
\addcontentsline{toc}{section}{References}

\bibliography{esaalburg}

\begin{thebibliography}{10}

\bibitem{schmitz}
N.~Schmitz, {\em Neutrinophysik\/} (Teubner Studienbücher, 1997).

\bibitem{raffelt}
G.~Raffelt, {\em Stars as Laboratories for Fundamental Physics\/} (Chicago
  Press, 1996).

\bibitem{kolbturner}
{E. W. Kolb, M.S. Turner}, {\em The Early Universe\/} (Addison Wesley, 1993).

\bibitem{zuber}
{H.V. Klapdor--Kleingrothaus, K. Zuber}, {\em Teilchenastrophysik\/} (Teubner
  Studienbücher, 1997).

\bibitem{bergstrom}
{L.~Bergstr\"om, A.~Goobar}, {\em Cosmology and Particle Astrophysics\/}
  (Wiley, 1999).

\bibitem{zuber2}
K.~Zuber, {\em Phys. Rep.\/} {\bf 305} (1998) 295.

\bibitem{raffelt98}
G.~Raffelt, hep-ph/9902271.

\bibitem{raffelt99}
G.~Raffelt, {\em Annu. Rev. Nucl. Part. Sci.\/} {\bf 49} (1999) in press.

\bibitem{bilenky99}
W.~G. S.~M.~Bilenky, C.~Giunti, {\em Prog. Part. Nucl. Phys.\/} {\bf 43} (1999)
  1.

\bibitem{clayton}
{D.~D.~Clayton}, {\em Principles of Stellar Evolution and Nucleosynthesis\/}
  (University of Chicago Press, 1968).

\bibitem{kippenhahn}
{R.~Kippenhahn, A.~Weigert}, {\em Stellar Structure and Evolution\/} (Springer,
  1990).

\bibitem{bahcall}
{J. N. Bahcall, S. Basu, M. H. Pinsonneault}, {\em Phys. Lett.\/} {\bf B 433}
  (1998) 1.

\bibitem{SKdata}
{(SuperKamiokande collaboration) Y.~ Fukuda}, {\em Phys. Rev. Lett.\/} {\bf 81}
  (1998) 1562.

\bibitem{IMB}
{(IMB collaboration) R.~ Becker--Szendy {\it et al.}}, {\em Nucl. Phys. {\bf B}
  Proc. Suppl\/} {\bf 83} (1995) 331.

\bibitem{MACRO}
{(MACRO collaboration) M.~ Ambrosio {\it et al.}}, {\em Phys. Lett.\/} {\bf B
  434} (1998) 451.

\bibitem{Soudan}
{(Soudan collaboration) W.~ W.~ M.~ Allison {\it et al.}}, {\em Phys. Lett.\/}
  {\bf B 449} (1999) 137.

\bibitem{LSND}
{(LSND collaboration) C. Athanassopoulos {\it et al.}}, {\em Phys. Rev.\/} {\bf
  C 58} (1998) 2489.

\bibitem{KARMEN}
{(KARMEN collaboration) B. Zeitnitz {\it et al.}}, {\em Prog. Nucl. Part.
  Phys.\/} {\bf 40} (1998) 169.

\bibitem{CHOOZ}
{(CHOOZ collaboration) M. Apollonio {\it et al.}}, {\em Phys. Lett.\/} {\bf B
  420} (1998) 397.

\bibitem{homestake}
{(Homestake Kollaboration) R. Davis {\it et al.}}, {\em Nucl. Phys. (Proc.
  Suppl.)\/} {\bf B 48} (1996) 284.

\bibitem{gallex}
{(GALLEX collaboration) W. Hampel {\it et al.}}, {\em Phys. Lett.\/} {\bf B
  388} (1996) 364.

\bibitem{sage}
(SAGE Kollaboration) V. Gavrin in {\it Proceedings of the XVIII International
  Conference on Neutrino Physics and Astrophysics}, Takayama, Japan, Juni 1998.

\bibitem{SKsol}
{(SuperKamiokande collaboration){\it et al.}}, {\em Phys. Rev. Lett.\/} {\bf
  81} (1998) 1158.

\bibitem{kamiokande}
{(Kamiokande collaboration) Y. Fukuda {\it et al.}}, {\em Phys. Rev. Lett.\/}
  {\bf 77} (1996) 1683.

\bibitem{solpar}
{J. N. Bahcall, P. I. Krastev, A. Y. Smirnov}, {\em Phys. Rev.\/} {\bf D 58}
  (1998) 096016.

\bibitem{SNO}
(SNO collaboration) J. Boger {\it et al.}, nucl-ex/9910016.

\bibitem{borex}
L.~Oberauer, {\em Nucl. Phys. B (Proc. Suppl.)\/} {\bf 77} (1999) 48.

\bibitem{miniboone}
E.~ Church {\it et al.} nucl-ex/9706011.

\bibitem{K2K}
K.~Nishikawa, {\em Nucl. Phys. (Proc. Suppl.)\/} {\bf B 77} (1999) 198.

\bibitem{MINOS}
S.~Wojcicki, {\em Nucl. Phys. (Proc. Suppl.)\/} {\bf B 77} (1999) 182.

\bibitem{ICARUS}
C.~Rubbia, {\em Nucl. Phys. (Proc. Suppl.)\/} {\bf B 48} (1996) 172.

\bibitem{BurTyt}
{S. Burles, D. Tytler}, {\em Astrophys. J.\/} {\bf 499} (1998) 699, \\ {\it
  ibid.} {\bf 507} 732.

\bibitem{Bur}
{S. Burles, {\it et al.}}, {\em Phys. Rev. Lett.\/} {\bf 82} (1999) 4176, in
  press.

\bibitem{TreGun}
{S. Tremaine, J. Gunn}, {\em Phys. Rev. Lett.\/} {\bf 42} (1979) 407.

\end{thebibliography}

\end{document}